\documentclass[11pt]{article}

\usepackage[T1]{fontenc}
\usepackage[utf8]{inputenc}
\usepackage{tgtermes}
\usepackage{amsmath}
\usepackage{amssymb}
\usepackage{bm}
\usepackage{graphicx}
\usepackage{booktabs}
\usepackage{tabularx}
\usepackage{chemformula}
\usepackage{microtype}
\usepackage{setspace}
\usepackage[super,sort&compress]{natbib}
\usepackage[colorlinks=true,allcolors=blue]{hyperref}
\usepackage[margin=1in]{geometry}

\newcommand{\Ik}{I_k}

\title{Crystal symmetry predicts unconventional magnetism}

\author{\parbox{0.97\textwidth}{\centering
\setstretch{1.1}\normalsize
Ziyin Song$^{1,2}$, Zhong Fang$^{1,2}$, Chen Fang$^{1,}$\thanks{Corresponding author: \href{mailto:cfang@iphy.ac.cn}{cfang@iphy.ac.cn}.}, and Hongming Weng$^{1,3,}$\thanks{Corresponding author: \href{mailto:hmweng@iphy.ac.cn}{hmweng@iphy.ac.cn}.}\\[0.8em]
\small
$^{1}$Beijing National Laboratory for Condensed Matter Physics,\\
and Institute of Physics, Chinese Academy of Sciences, Beijing 100190, China\\[0.35em]
$^{2}$University of Chinese Academy of Sciences, Beijing 100049, China\\[0.35em]
$^{3}$Condensed Matter Physics Data Center, Chinese Academy of Sciences, Beijing 100190, China
}}
\date{}

\begin{document}

\maketitle

\begin{abstract}
Unconventional compensated magnets combine zero net magnetization with momentum-dependent spin polarization, but identifying them usually requires knowledge of their magnetic order. Here we show that crystal symmetry can constrain unconventional magnetic character before the magnetic ground state is known. Starting from a non-magnetic crystal structure and a specified magnetic sublattice, we generate symmetry-compatible compensated orders and classify their spin textures using spin-space-group symmetry. We identify materials whose generated candidates are all unconventional, either across a defined search space or after restricting the magnetic-cell size. Within the experimental benchmark, 68\% of the prioritized materials are unconventional, compared with 9\% of the remaining materials. Screening the Materials Project yields thousands of promising candidates for unconventional compensated magnetism. First-principles calculations for \ch{VGe3} and tetragonal \ch{Fe2SiO4} connect these symmetry predictions to the energetics and spin textures of competing magnetic orders. In \ch{VGe3}, a noncoplanar candidate permits mixed-wave spin polarization along a fixed axis without spin--orbit coupling, combining components that are odd and even under momentum reversal. This framework enables crystallography-guided searches for unconventional compensated magnets without first determining their magnetic ground states.
\end{abstract}

% Literature citations will be inserted after bibliography curation.

% opening, can without section name

Zero net magnetization does not imply a spin-degenerate electronic structure. Unconventional compensated magnets, including altermagnets\cite{smejkal2021altermagnetism,TJungwirth2022} and $p$-wave magnets\cite{hellenes2024pwavemagnets,brekke2024minimal}, can exhibit momentum-dependent spin polarization even without spin--orbit coupling\cite{Noda2016altermagnetism,Ahn2019altermagnetism,Yuan2020altermagnetism}. Their spin textures open routes to spin-polarized transport\cite{brekke2024minimal,Chakraborty2025} and unconventional Hall responses\cite{smj2020altermagnetism,liu2025multipolar}. Identifying such materials, however, usually starts from a known magnetic structure: the arrangement of local moments determines the symmetries that constrain the electronic spin texture.

Complete magnetic structures are experimentally demanding to resolve and are known for far fewer compounds than crystallographic structures. This disparity limits searches based on established magnetic orders, leaving a much larger pool of structurally characterized materials to explore. Yet establishing a material's broad magnetic character may not require identifying its precise magnetic order. Can crystal symmetry constrain whether a magnet is unconventional before the order it adopts is known?

Here we address this question through properties shared across an ensemble of candidate magnetic orders. Spin-space-group (SSG) symmetry connects real-space magnetic configurations to reciprocal-space spin textures through combined transformations in real and spin space\cite{Brinkman1966,Litvin1974,liu2022SSG}. Starting from a non-magnetic crystal structure and a specified magnetic sublattice, we generate symmetry-compatible compensated orders within a defined search space and classify their spin textures. If every candidate is unconventional, this character is independent of energetic selection within the ensemble. Applying this criterion across different magnetic-cell sizes establishes a hierarchy of crystallographic constraints on unconventional magnetism.

Benchmarking against experimentally established magnetic structures shows that unconventional order is substantially more common among materials prioritized by the hierarchy. Applying the framework to the Materials Project identifies thousands of promising candidates for unconventional compensated magnetism. First-principles studies of \ch{VGe3} and tetragonal \ch{Fe2SiO4} connect these predictions to the energetics and spin textures of competing magnetic orders. A noncoplanar candidate of \ch{VGe3} further illustrates mixed-wave spin polarization, in which components even and odd under momentum reversal coexist along a fixed spin axis without spin--orbit coupling.

\section*{Symmetry predicts unconventional magnetism}

The framework takes a non-magnetic crystal structure and a specified magnetic sublattice as inputs. We consider one magnetic species occupying a single symmetry-equivalent Wyckoff position and retain candidates with magnetic-cell index $I_k\leq4$. Here, $I_k$ measures the enlargement of the magnetic primitive cell relative to the crystallographic primitive cell. The spatial projection of each candidate SSG preserves the parent crystallographic symmetry. These conditions define the search space within which we predict compensated magnetic character.

\begin{figure*}[t]
    \centering
    \includegraphics[width=\textwidth]{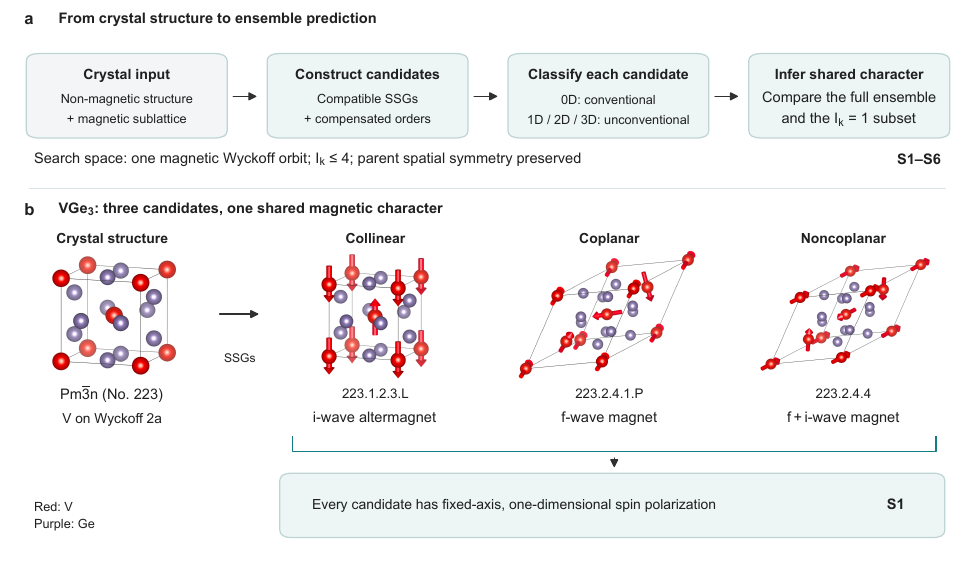}
    \caption{\textbf{Symmetry-based prediction of unconventional compensated magnetism.}
    \textbf{a}, A non-magnetic crystal structure and its magnetic sublattice define compatible spin-space groups and compensated magnetic configurations within the chosen symmetry search space. Each candidate is classified by its reciprocal-space spin-texture dimensionality. Properties shared by the candidate ensemble, with magnetic-cell information where available, determine the material-level classes S1--S6.
    \textbf{b}, \ch{VGe3} illustrates the S1 class: its collinear $i$-wave, coplanar $f$-wave and noncoplanar $f+i$-wave candidates all permit spin polarization along a fixed axis. Red and purple spheres denote V and Ge, respectively; arrows show magnetic moments.}
    \label{fig:ssg_prediction_workflow}
\end{figure*}

For each parent structure, we enumerate compatible SSGs\cite{xiao2023spin,jiang2023enumeration,ren2023enumeration} and construct their magnetic configurations on the specified sublattice (Fig.~\ref{fig:ssg_prediction_workflow}). The SSG operations constrain both the arrangement of moments on symmetry-related sites and the associated reciprocal-space spin texture. We retain configurations with vanishing net magnetization whose full magnetic symmetry matches the target SSG. Each material therefore yields an ensemble of compatible compensated candidates, as detailed in Supplementary Information Section~5.

Each candidate is characterized in real and reciprocal space. The real-space magnetic configuration is collinear, coplanar or noncoplanar, whereas its symmetry-allowed spin texture is zero-, one-, two- or three-dimensional\cite{SongUnifiedPRX,Luo2026Unconventional,Lange2026AntiSpinLaue}. Zero-dimensional textures identify conventional compensated states; finite-dimensional textures define unconventional magnetism in the present classification.

For one-dimensional, spin-polarized textures, the polarization lies along a fixed spin-space axis. Its scalar momentum dependence is classified using the lowest-order symmetry-allowed polynomial basis functions\cite{SongUnifiedPRX,oddSSG}. The conserved hyperspin component\cite{Ma2026HyperspinAltermagnets} is the normalized operator representation of the spin-translation symmetry that fixes the polarization axis. This protection does not require $\Theta T_{\boldsymbol\tau}$ (Supplementary Information Section~4). The resulting wave symmetry can be odd, such as $p$ or $f$, even, such as $d$ or $g$, or mixed. In mixed-wave states, odd and even components contribute along the same polarization axis. Their lowest allowed orders define an additive label, such as $p+d$, with material-dependent amplitudes. Related spin-splitting and transport phenomena have been studied in collinear spin-orbital magnets and noncoplanar \ch{CrSe}\cite{Zhuang2026MixedParityCollinear,Shen2026NiAsMixedParity,Zhu2025MagneticGeometry}. Here, noncoplanar magnetic order provides a route in which SSG symmetry fixes the polarization axis while allowing both odd and even momentum dependences. The symmetry origin and a material example are discussed in Supplementary Information Section~10.

In a specified spin coordinate system, higher-dimensional textures can exhibit different wave symmetries across components, termed hybrid-wave\cite{Luo2026Unconventional,Ryu2026MixedParityTetraborides,Hu2026NiS2}. Their component basis functions satisfy the full vector symmetry constraints, as detailed in Supplementary Information Section~3.

We organize materials into six classes, S1--S6, according to the selectivity of their candidate ensembles. In S1, every candidate with $I_k\leq4$ is a spin-polarized one-dimensional state. In S2, every candidate is unconventional, although higher-dimensional textures may also occur. For materials outside S1 and S2, S3 and S4 contain mixed full ensembles but exclusively spin-polarized or unconventional $I_k=1$ subsets, respectively. Their selectivity therefore depends on knowing that the magnetic order does not enlarge the crystallographic primitive cell. S5 retains both conventional and unconventional possibilities, whereas S6 contains only conventional candidates. The formal definitions appear in Supplementary Information Section~6. These classes express symmetry constraints within a specified candidate space, rather than statistical probabilities.

\section*{Experimental structures validate the predictions}

We benchmark the predictions against MAGNDATA\cite{Gallego2016magndataI,Gallego2016magndataII}, which contains experimentally refined magnetic structures. We remove the experimental spin configuration from each input and regenerate candidates using only the non-magnetic crystallographic structure and magnetic-sublattice information. Independently classifying the experimental SSG by spin-texture dimensionality then tests whether the ensemble predicts its magnetic character.

After structural filtering, 862 zero-net-moment antiferromagnetic entries admit at least one compatible candidate; Supplementary Information Section~7 gives the filtering statistics and exclusions. The generated ensembles contain the experimentally reported SSG for 571 entries. Among 380 entries with experimental magnetic-cell index $I_k=1$, 348 ensembles contain the reported SSG (91.58\%). The search space thus includes many experimentally realized orders, although exact structure recovery is not the primary prediction target.

The benchmark also tests whether ensemble selectivity tracks the magnetic character of the experimental state. Of the five S1 materials, three have spin-polarized experimental states. In S2, 43 of 52 materials are unconventional, whereas 43 of 45 S6 materials are conventional. The magnetic-cell-conditioned classes show a similar separation: 64 of 128 S3 materials are spin polarized, and 70 of 93 S4 materials are unconventional. In the mixed S5 class, 490 of 539 materials are conventional. These results support selective prioritization, while the mismatches show that ensemble predictions do not guarantee the experimental character outside the assumed search space.

Knowledge of the magnetic-cell size sharpens this test without requiring the complete spin configuration. Among benchmark entries with experimental $I_k=1$, we identify 161 whose generated $I_k=1$ candidates are all unconventional. All 161 also have unconventional experimental states. This agreement shows how crystallographic and magnetic-cell information can identify unconventional character within the benchmark, even when the detailed magnetic order is withheld.

\section*{Large-scale discovery of unconventional magnets}

We next apply the framework to the Materials Project\cite{Jain2013MaterialsProject,Horton2025MaterialsProject}, where crystallographic structures are available for many compounds without experimentally resolved magnetic order. Of 58,996 magnetic entries, 23,428 satisfy the structural conditions and admit at least one compensated candidate with $I_k\leq4$. Together, they generate 825,243 compatible magnetic candidates (Supplementary Information Section~8).

Among the retained materials, 1,769 belong to S1 or S2, so every candidate in their full $I_k\leq4$ ensemble is unconventional. Of these, 533 have exclusively spin-polarized candidates. Restricting the candidate space to $I_k=1$ identifies 6,248 materials with only unconventional candidates, including 3,345 with only spin-polarized candidates. By contrast, 14,329 materials belong to mixed S5 ensembles and 2,537 to conventional-only S6 ensembles. The search thus distinguishes materials with strongly constrained unconventional character from those that retain conventional possibilities.

To highlight materials suited to the weak-SOC regime of SSG symmetry\cite{liu2022SSG}, the accompanying repository includes a complete catalogue of S1--S4 entries satisfying a light-element composition filter (Supplementary Information Section~8).

The selected ensembles provide targets for diffraction, spectroscopy and first-principles calculations. Their predictive value lies in the character shared by all candidates within the relevant search space, which can be assessed before detailed magnetic characterization. Energetic calculations can then prioritize configurations within these ensembles, as illustrated by the following material examples.

\section*{From symmetry-selected ensembles to electronic structure}

\begin{figure*}[p]
    \centering
    \includegraphics[width=\textwidth]{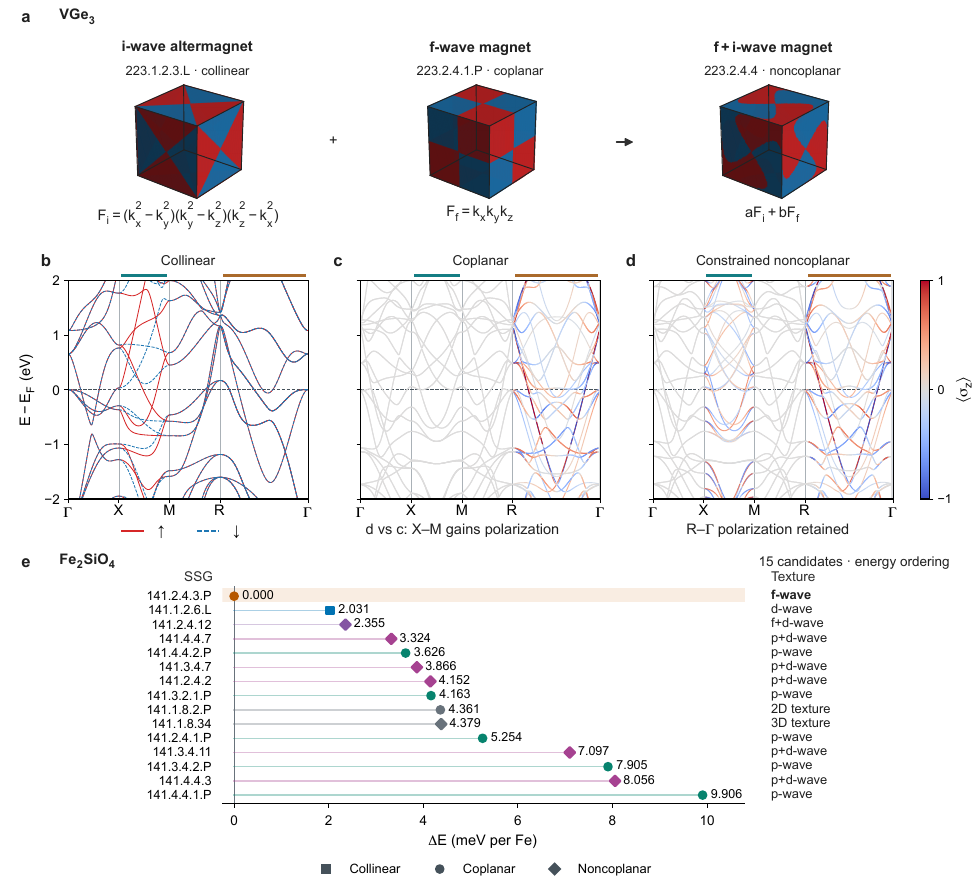}
    \begingroup
    \setstretch{1.0}\small
    \caption{\textbf{Symmetry-selected magnetic states in \ch{VGe3} and \ch{Fe2SiO4}.}
    \textbf{a}, Momentum-space basis functions for the three \ch{VGe3} candidates: the collinear $i$-wave altermagnet (SSG 223.1.2.3.L), coplanar $f$-wave magnet (223.2.4.1.P), and noncoplanar $f+i$-wave magnet (223.2.4.4). Red and blue indicate positive and negative values on a common momentum-space cube. The combination $aF_i(\mathbf{k})+bF_f(\mathbf{k})$ illustrates the symmetry-allowed coexistence of even $i$-wave and odd $f$-wave components; the relative coefficients are schematic.
    \textbf{b--d}, Corresponding DFT+$U$ band structures ($U=3$~eV) without spin--orbit coupling, along the same physical momentum path represented in each magnetic cell. Panel \textbf{b} resolves the two spin channels; panels \textbf{c} and \textbf{d} are coloured by the projected spin polarization $\langle\sigma_z\rangle$, averaged within near-degenerate subspaces. Panel \textbf{d} uses a constrained noncoplanar configuration canted by approximately $10^\circ$ from the coplanar state. Teal and ochre bars mark the $X$--$M$ and $R$--$\Gamma$ intervals, respectively, highlighting the additional and retained polarization in \textbf{d} relative to \textbf{c}.
    \textbf{e}, Relative DFT+$U$ energies ($U=4$~eV) of all 15 \ch{Fe2SiO4} candidates, ordered by energy, normalized per Fe atom and referenced to the lowest-energy candidate, the coplanar $f$-wave state with SSG 141.2.4.3.P (shaded row). Colours identify the symmetry-derived texture types; squares, circles and diamonds denote collinear, coplanar and noncoplanar magnetic configurations, respectively. Computational details are given in Supplementary Information, Section~11. Numerical SSG labels are matched to international symbols in Supplementary Table~\ref{tab:ssg_international_symbols}.}
    \label{fig:vge3_case}
    \endgroup
\end{figure*}

The S1 and S2 classes constrain magnetic character at different levels, as illustrated by \ch{VGe3} and tetragonal \ch{Fe2SiO4}, respectively (Fig.~\ref{fig:vge3_case}). In \ch{VGe3}, a compact ensemble connects three distinct real-space magnetic geometries to one-dimensional spin polarization. In \ch{Fe2SiO4}, a larger ensemble includes both one- and higher-dimensional textures, allowing energetic calculations to prioritize a particular state within an entirely unconventional candidate set.

For \ch{VGe3} (mp-672337), the parent structure has space group $Pm\bar{3}n$ (No.~223), with V occupying a single symmetry-equivalent Wyckoff position. The $I_k\leq4$ search yields only three compensated candidates: a collinear $i$-wave altermagnet with SSG 223.1.2.3.L~[$\mathrm{P}^{1,1,1}\,\mathrm{m}^{1}\,\bar{3}^{1}\,\mathrm{n}^{\bar{1}}\,\left(C_{i}^{\mathrm{I}}\right)$] and $I_k=1$, and two $I_k=2$ states, the coplanar $f$-wave magnet 223.2.4.1.P~[$\mathrm{P}^{2,2,2}\,\mathrm{m}^{m}\,\bar{3}^{m}\,\mathrm{n}^{4}\,\left(C_{4v}^{\mathrm{II}}\right)$] and noncoplanar $f+i$-wave magnet 223.2.4.4~[$\mathrm{P}^{2,2,2}\,\mathrm{m}^{m}\,\bar{3}^{m}\,\mathrm{n}^{\bar{4}^{-1}}\,\left(D_{2d}^{\mathrm{III}}\right)$]. The international symbols follow the IRSSG convention\cite{Zhang2025IRSSG}. All three have one-dimensional reciprocal-space spin textures, placing the material in S1 independently of their energetic ordering within this search space.

\begingroup\clubpenalty=10000
First-principles calculations favour the coplanar $f$-wave solution over the collinear $i$-wave state, while an unconstrained noncoplanar initialization evolves towards the coplanar solution (Supplementary Information, Section~9). The band structures reveal how the three candidates differ electronically (Fig.~\ref{fig:vge3_case}b--d). The collinear state exhibits splitting between two conserved spin channels, whereas the coplanar state has a non-quantized, momentum-dependent spin projection along a fixed axis\cite{Ma2026HyperspinAltermagnets}. Along the plotted path, the constrained noncoplanar configuration combines features of both limits: it retains the $f$-wave polarization on $R$--$\Gamma$ and develops additional spin-polarized splitting on $X$--$M$, where symmetry enforces zero spin polarization in the coplanar state. The mixed-wave state thus brings together the characteristic momentum dependences of the altermagnetic and coplanar states while preserving one-dimensional spin polarization.\par
\endgroup

This connection is also reflected in the momentum-space sign patterns (Fig.~\ref{fig:vge3_case}a). The odd $f$-wave and even $i$-wave components contribute to the same scalar polarization, whose combined pattern generally has no definite parity. The intersections of their nodal planes through $\Gamma$, where both components vanish, remain exact nodal lines protected by SSG operations. Continuity also requires zero-polarization boundaries between regions of opposite sign, which generically form nodal surfaces\cite{Urru2025BiFeO3NodalSurfaces}. These surfaces can bend as the relative contributions change, while the symmetry-protected lines remain fixed. Nodal lines and nodal surfaces in the mixed-wave state are discussed in Supplementary Information, Section~10.

\begingroup\emergencystretch=1em
Tetragonal \ch{Fe2SiO4} (mp-18816), with parent space group $I4_1/amd$ (No.~141), extends this picture to a larger S2 ensemble. Its 15 candidates consist of one $d$-wave altermagnet, one two-dimensional texture, one three-dimensional texture and six pairs of coplanar odd-wave and noncoplanar mixed-wave states. In each pair, adding the staggered moment pattern of the $d$-wave altermagnet perpendicular to the coplanar spin plane produces the mixed-wave candidate: five pairs connect $p$-wave to $p+d$-wave order, and one connects $f$-wave to $f+d$-wave order. This repeated pairing extends the relation illustrated by \ch{VGe3} to a richer set of magnetic configurations. The energy comparison identifies the coplanar $f$-wave state 141.2.4.3.P~[$\mathrm{I}^{2,2,2}\,4_1^{m}/\,\mathrm{a}^{m}\,\mathrm{m}^{m}\,\mathrm{d}^{4^{-1}}\,\left(C_{4v}^{\mathrm{II}}\right)$] as the lowest-energy candidate (Fig.~\ref{fig:vge3_case}e), selecting a one-dimensional spin-polarized state from an ensemble that also admits higher-dimensional textures.\par
\endgroup

These examples illustrate the complementary roles of symmetry and energetics. The candidate ensemble establishes the shared unconventional character before the magnetic order is known, while energy comparisons identify the preferred configuration and its associated electronic signatures.

% ending paragraph:

\begin{samepage}
Taken together, these results establish a route to predicting unconventional magnetic character from crystallographic information before the realized magnetic order is known. The predictive information lies in properties shared across the symmetry-compatible candidate ensemble, allowing materials to be prioritized without first identifying a unique magnetic ground state. This symmetry-based selectivity provides a focused starting point for microscopic energetic calculations\cite{Li2025MagneticGroundStates,Zhou2026MagneticStructuresDatabase} and experimental characterization. Extending the search to larger magnetic cells, magnetic orders with reduced spatial symmetry relative to the parent crystal, and multiple magnetic Wyckoff orbits would broaden the accessible materials space. The candidate-ensemble approach thus turns readily available crystallographic information into a practical guide for discovering unconventional compensated magnets.\par
\end{samepage}

% Planned next sections:
% \section*{Symmetry generates compensated magnetic order}
% \section*{Experimental structures validate the prediction}
% \section*{A large space of unconventional magnets}
% \section*{Energetics selects a spin-polarized state}

\section*{Methods}

\subsection*{Construction of symmetry-compatible magnetic candidates}

The input is a standardized crystal structure with space group $G$ and one magnetic species occupying a single symmetry-equivalent Wyckoff orbit. We select candidate SSGs whose spatial projection equals $G$, whose magnetic-cell index satisfies $I_k\leq4$, and whose spin-space symmetry forbids a net magnetization. The last condition requires that no nonzero vector be invariant under every spin part of the SSG. These restrictions define the search space, including its magnetic-cell cutoff and retention of the full crystallographic spatial symmetry.

For each candidate, we construct the magnetic cell and select a reference site $\mathbf r_0$ on the magnetic orbit. Onsite operations $\{U_\alpha\Vert R_\alpha|\boldsymbol{\tau}_\alpha\}$ leave this site invariant modulo a magnetic lattice translation. The allowed reference moment satisfies
\begin{equation}
    U_\alpha\mathbf m_0=\mathbf m_0
\end{equation}
for every onsite operation. We obtain its allowed subspace from the null space of the stacked matrices $U_\alpha-I_3$, where $I_3$ is the identity matrix. A zero-dimensional subspace excludes a nonzero moment and therefore rules out the candidate. Otherwise, we choose a normalized representative within the allowed subspace, using a generic linear combination when more than one basis vector is available.

The remaining moments follow by applying the SSG operations to the reference site and moment,
\begin{equation}
    \mathbf r_i=R_\alpha\mathbf r_0+\boldsymbol{\tau}_\alpha
    \pmod{\text{magnetic lattice}},
    \qquad
    \mathbf m_i=U_\alpha\mathbf m_0.
\end{equation}
Because the magnetic sites form one symmetry orbit, this construction determines their relative moment directions and equal magnitudes from a single reference moment.

We then determine the symmetry operations of the generated configuration and compare its parent space group, operation counts and real-space spin dimension with the target. The target operations are imposed by construction; this check rejects representatives with accidental additional symmetry. Only configurations passing this verification enter the candidate ensemble. The complete construction and material-level aggregation are described in Supplementary Information Sections~5 and~6.

\subsection*{Classification of reciprocal-space spin textures}

We classify each retained SSG by the spin polarization allowed at generic momentum. For an operation $g=\{U\Vert R|\boldsymbol{\tau}\}$, the symmetry constraint in Cartesian momentum coordinates is
\begin{equation}
    \mathbf S(\mathbf k)
    =U\,\mathbf S\!\left[\det(U)R^{-1}\mathbf k\right].
    \label{eq:method_kspace_transform}
\end{equation}
Here $U$ acts in spin space and $R$ is the spatial point operation. The factor $\det(U)=-1$ supplies the additional momentum reversal for antiunitary operations; $\det(U)=+1$ applies to unitary operations. The translation $\boldsymbol{\tau}$ does not shift momentum in this relation for the spin expectation value.

Operations that leave a generic $\mathbf k$ invariant impose $U_\alpha\mathbf S(\mathbf k)=\mathbf S(\mathbf k)$. We compute their common invariant subspace and its dimension,
\begin{equation}
    V_{\mathbf k}=\bigcap_\alpha\mathrm{Fix}(U_\alpha),
    \qquad
    d=\dim V_{\mathbf k}\in\{0,1,2,3\}.
    \label{eq:method_texture_dim}
\end{equation}
We classify $d=0$ states as conventional compensated magnets and $d>0$ states as unconventional. For $d=1$, the polarization has the form $\mathbf S(\mathbf k)=s(\mathbf k)\hat{\mathbf n}$, with a fixed spin-space axis $\hat{\mathbf n}$. States with $d=2$ or $d=3$ allow polarization within a plane or the full spin space, respectively. These dimensions specify the symmetry-allowed subspace, rather than requiring nonzero polarization at every momentum.

For one-dimensional textures, we solve the homogeneous-polynomial constraints separately in the odd and even momentum-parity sectors. Their lowest allowed degrees determine the wave labels; the noncoplanar spin-1D classification examines polynomials through degree nine. When both sectors are allowed, we combine their leading labels in odd--even order, such as $f+i$, to denote mixed-wave symmetry. The label describes allowed basis functions for the same scalar polarization, while their amplitudes depend on the electronic structure. Basis functions and classification details are given in Supplementary Information Section~3. We evaluate the classification once per SSG and assign it to every material-specific candidate realizing that symmetry.

\subsection*{First-principles calculations}

Electronic-structure calculations used VASP\cite{Kresse1996VASP}, the projector-augmented-wave method\cite{Blochl1994PAW,Kresse1999PAW} and the Perdew--Burke--Ernzerhof (PBE) exchange--correlation functional\cite{Perdew1996PBE}. Transition-metal $3d$ correlations were treated with the rotationally invariant DFT+$U$ method\cite{Dudarev1998DFTU}, using $J=0$ and the material-specific $U$ values below. Spin--orbit coupling was omitted to examine the non-relativistic states described by SSG symmetry. All calculations kept the lattice vectors and atomic coordinates fixed, and energy comparisons within each material used identical PAW datasets. Gaussian smearing with $\sigma=0.05$~eV was used for both materials.

\paragraph{\ch{VGe3}.}
Calculations used $U=3$~eV on V $3d$ states, a 500~eV plane-wave cutoff and electronic self-consistency thresholds of $10^{-6}$--$10^{-7}$~eV. Self-consistent calculations used $\Gamma$-centred $9\times9\times9$ meshes in the corresponding magnetic cells. The $223.1.2.3.\mathrm{L}$ candidate was treated with collinear spin polarization; the $223.2.4.1.\mathrm{P}$ and $223.2.4.4$ candidates used noncollinear calculations. For energy comparisons, the $I_k=1$ collinear pattern was also represented in a doubled, 16-atom cell matching the $I_k=2$ candidates. Relative energies were obtained from $E_{\sigma\rightarrow0}$ and normalized per formula unit.

Band structures were calculated non-self-consistently from the corresponding self-consistent charge densities and referenced to their Fermi energies. The collinear bands used an eight-atom cell; the coplanar and constrained noncoplanar bands used 16-atom cells. We expressed the same physical momentum path in each reciprocal basis, without band unfolding. Collinear bands resolve the two spin channels, whereas noncollinear bands in Fig.~\ref{fig:vge3_case} are coloured by $\langle\sigma_z\rangle$. For the latter, states within 1~meV were grouped at each momentum. Each group was assigned the ratio of its summed \texttt{PROCAR} spin-$z$ projection to its summed charge weight.

The constrained noncoplanar calculation started from the coplanar charge density, with V moments canted by approximately $10^\circ$ towards the noncoplanar directions. Local-moment directions were constrained using \texttt{I\_CONSTRAINED\_M=1} and \texttt{LAMBDA=10}, while their magnitudes remained free. This calculation probes the noncoplanar electronic structure and is excluded from the unconstrained energy comparison. Initial moment patterns, energy comparisons and momentum-path conventions are documented in Supplementary Information Section~9.

\paragraph{Tetragonal \ch{Fe2SiO4}.}
The 15 symmetry-generated candidates were screened using static, unconstrained noncollinear PBE+$U$ calculations with $U=4$~eV on Fe $3d$ states and a 600~eV cutoff. The $I_k=1,2$ candidates shared a 28-atom cell with a $\Gamma$-centred $9\times9\times6$ mesh. The $I_k=3$ and $I_k=4$ candidates used 42- and 56-atom cells with $9\times9\times4$ and $9\times9\times3$ meshes, respectively. Final electronic-step energy changes were below $10^{-5}$~eV per Fe atom. Energies $E_{\sigma\rightarrow0}$ were normalized per Fe atom and referenced to the lowest-energy candidate, the coplanar $f$-wave state. The candidate configurations and relative energies are given in Supplementary Information Section~11.

\section*{Code availability}

The code, Materials Project material catalogue and input files for the first-principles calculations are available on the MatElab platform at \url{https://eln.iphy.ac.cn/eln/link.html\#/208/unconv}.

\section*{Acknowledgements}

This work was supported by the Science Center of the National Natural Science Foundation of China (Grants No. 12188101 and No. 12325404), the National Key R\&D Program of China (Grants No. 2022YFA1403800, No. 2023YFA1607400, No. 2024YFA1408400 and No. 2023YFA1406704), the National Natural Science Foundation of China (Grants No. 12274436 and No. 12547112), the Strategic Priority Research Program (B) of the Chinese Academy of Sciences (CAS) (Grant No. XDB1720000), and H. W. acknowledges support from the New Cornerstone Science Foundation through the XPLORER PRIZE.

The governance of data was performed on the MatElab platform, developed by the Condensed Matter Physics Data Center of Chinese Academy of Sciences.

\clearpage
% A single reference list is shared by the main text and the supplement.
\bibliographystyle{unsrtnat}
\bibliography{unconv_prediction}

@article{xiao2023spin,
  title = {Spin Space Groups: Full Classification and Applications},
  author = {Xiao, Zhenyu and Zhao, Jianzhou and Li, Yanqi and Shindou, Ryuichi and Song, Zhi-Da},
  journal = {Phys. Rev. X},
  volume = {14},
  issue = {3},
  pages = {031037},
  numpages = {33},
  year = {2024},
  month = {Aug},
  publisher = {American Physical Society},
  doi = {10.1103/PhysRevX.14.031037},
  url = {https://link.aps.org/doi/10.1103/PhysRevX.14.031037}
}

@article{ren2023enumeration,
  title = {Enumeration and Representation Theory of Spin Space Groups},
  author = {Chen, Xiaobing and Ren, Jun and Zhu, Yanzhou and Yu, Yutong and Zhang, Ao and Liu, Pengfei and Li, Jiayu and Liu, Yuntian and Li, Caiheng and Liu, Qihang},
  journal = {Phys. Rev. X},
  volume = {14},
  issue = {3},
  pages = {031038},
  numpages = {33},
  year = {2024},
  month = {Aug},
  publisher = {American Physical Society},
  doi = {10.1103/PhysRevX.14.031038},
  url = {https://link.aps.org/doi/10.1103/PhysRevX.14.031038}
}

@article{jiang2023enumeration,
  title = {Enumeration of Spin-Space Groups: Toward a Complete Description of Symmetries of Magnetic Orders},
  author = {Jiang, Yi and Song, Ziyin and Zhu, Tiannian and Fang, Zhong and Weng, Hongming and Liu, Zheng-Xin and Yang, Jian and Fang, Chen},
  journal = {Phys. Rev. X},
  volume = {14},
  issue = {3},
  pages = {031039},
  numpages = {25},
  year = {2024},
  month = {Aug},
  publisher = {American Physical Society},
  doi = {10.1103/PhysRevX.14.031039},
  url = {https://link.aps.org/doi/10.1103/PhysRevX.14.031039}
}

@article{liu2025multipolar,
  title = {Multipolar Anisotropy in Anomalous Hall Effect from Spin-Group Symmetry Breaking},
  author = {Liu, Zheng and Wei, Mengjie and Peng, Wenzhi and Hou, Dazhi and Gao, Yang and Niu, Qian},
  journal = {Phys. Rev. X},
  volume = {15},
  issue = {3},
  pages = {031006},
  numpages = {23},
  year = {2025},
  month = {Jul},
  publisher = {American Physical Society},
  doi = {10.1103/PhysRevX.15.031006},
  url = {https://link.aps.org/doi/10.1103/PhysRevX.15.031006}
}

@article{Litvin1974,
title = {Spin groups},
journal = {Physica},
volume = {76},
number = {3},
pages = {538-554},
year = {1974},
issn = {0031-8914},
doi = {https://doi.org/10.1016/0031-8914(74)90157-8},
url = {https://www.sciencedirect.com/science/article/pii/0031891474901578},
author = {D.B. Litvin and W. Opechowski}
}

@article{Brinkman1966,
 ISSN = {00804630},
 url = {http://www.jstor.org/stable/2415409},
 author = {W. F. Brinkman and R. J. Elliott},
 journal = {Proceedings of the Royal Society of London. Series A, Mathematical and Physical Sciences},
 number = {1438},
 pages = {343--358},
 publisher = {The Royal Society},
 title = {Theory of Spin-Space Groups},
 volume = {294},
 year = {1966}
}

@article{liu2022SSG,
  title = {Spin-Group Symmetry in Magnetic Materials with Negligible Spin-Orbit Coupling},
  author = {Liu, Pengfei and Li, Jiayu and Han, Jingzhi and Wan, Xiangang and Liu, Qihang},
  journal = {Phys. Rev. X},
  volume = {12},
  issue = {2},
  pages = {021016},
  numpages = {19},
  year = {2022},
  month = {Apr},
  publisher = {American Physical Society},
  doi = {10.1103/PhysRevX.12.021016},
  url = {https://link.aps.org/doi/10.1103/PhysRevX.12.021016}
}

@article{smejkal2021altermagnetism,
  title = {Beyond Conventional Ferromagnetism and Antiferromagnetism: A Phase with Nonrelativistic Spin and Crystal Rotation Symmetry},
  author = {\ifmmode \check{S}\else \v{S}\fi{}mejkal, Libor and Sinova, Jairo and Jungwirth, Tomas},
  journal = {Phys. Rev. X},
  volume = {12},
  issue = {3},
  pages = {031042},
  numpages = {16},
  year = {2022},
  month = {Sep},
  publisher = {American Physical Society},
  doi = {10.1103/PhysRevX.12.031042},
  url = {https://link.aps.org/doi/10.1103/PhysRevX.12.031042}
}

@article{TJungwirth2022,
  title = {Emerging Research Landscape of Altermagnetism},
  author = {\ifmmode \check{S}\else \v{S}\fi{}mejkal, Libor and Sinova, Jairo and Jungwirth, Tomas},
  journal = {Phys. Rev. X},
  volume = {12},
  issue = {4},
  pages = {040501},
  numpages = {27},
  year = {2022},
  month = {Dec},
  publisher = {American Physical Society},
  doi = {10.1103/PhysRevX.12.040501},
  url = {https://link.aps.org/doi/10.1103/PhysRevX.12.040501}
}

@article{Noda2016altermagnetism,
author = {Noda, Yusuke and Ohno, Kaoru and Nakamura, Shinichiro},
title  = {Momentum-dependent band spin splitting in semiconducting MnO2: a density functional calculation},
journal  = {Phys. Chem. Chem. Phys.},
year  = {2016},
volume  = {18},
issue  = {19},
pages  = {13294-13303},
publisher  = {The Royal Society of Chemistry},
doi  = {10.1039/C5CP07806G},
url  = {http://dx.doi.org/10.1039/C5CP07806G},
}

@article{smj2020altermagnetism,
      author = {Libor Šmejkal  and Rafael González-Hernández  and T. Jungwirth  and J. Sinova },
      title = {Crystal time-reversal symmetry breaking and spontaneous Hall effect in collinear antiferromagnets},
      journal = {Science Advances},
      volume = {6},
      number = {23},
      pages = {eaaz8809},
      year = {2020},
      doi = {10.1126/sciadv.aaz8809},
      URL = {https://www.science.org/doi/abs/10.1126/sciadv.aaz8809}
}

@article{Ahn2019altermagnetism,
  title = {Antiferromagnetism in ${\mathrm{RuO}}_{2}$ as $d$-wave Pomeranchuk instability},
  author = {Ahn, Kyo-Hoon and Hariki, Atsushi and Lee, Kwan-Woo and Kune\ifmmode \check{s}\else \v{s}\fi{}, Jan},
  journal = {Phys. Rev. B},
  volume = {99},
  issue = {18},
  pages = {184432},
  numpages = {5},
  year = {2019},
  month = {May},
  publisher = {American Physical Society},
  doi = {10.1103/PhysRevB.99.184432},
  url = {https://link.aps.org/doi/10.1103/PhysRevB.99.184432}
}

@article{Yuan2020altermagnetism,
  title = {Giant momentum-dependent spin splitting in centrosymmetric low-$Z$ antiferromagnets},
  author = {Yuan, Lin-Ding and Wang, Zhi and Luo, Jun-Wei and Rashba, Emmanuel I. and Zunger, Alex},
  journal = {Phys. Rev. B},
  volume = {102},
  issue = {1},
  pages = {014422},
  numpages = {13},
  year = {2020},
  month = {Jul},
  publisher = {American Physical Society},
  doi = {10.1103/PhysRevB.102.014422},
  url = {https://link.aps.org/doi/10.1103/PhysRevB.102.014422}
}

@misc{hellenes2024pwavemagnets,
      title={P-wave magnets}, 
      author={Anna Birk Hellenes and Tomáš Jungwirth and Rodrigo Jaeschke-Ubiergo and Atasi Chakraborty and Jairo Sinova and Libor Šmejkal},
      year={2024},
      eprint={2309.01607},
      archivePrefix={arXiv},
      primaryClass={cond-mat.mes-hall},
      url={https://arxiv.org/abs/2309.01607}, 
}

@article{brekke2024minimal,
  title = {Minimal Models and Transport Properties of Unconventional $p$-Wave Magnets},
  author = {Brekke, Bj\o{}rnulf and Sukhachov, Pavlo and Giil, Hans Gl\o{}ckner and Brataas, Arne and Linder, Jacob},
  journal = {Phys. Rev. Lett.},
  volume = {133},
  issue = {23},
  pages = {236703},
  numpages = {9},
  year = {2024},
  month = {Dec},
  publisher = {American Physical Society},
  doi = {10.1103/PhysRevLett.133.236703},
  url = {https://link.aps.org/doi/10.1103/PhysRevLett.133.236703}
}

@Article{Chakraborty2025,
author={Chakraborty, Atasi
and Birk Hellenes, Anna
and Jaeschke-Ubiergo, Rodrigo
and Jungwirth, Tom{\'a}s
and {\v{S}}mejkal, Libor
and Sinova, Jairo},
title={Highly efficient non-relativistic Edelstein effect in nodal p-wave magnets},
journal={Nature Communications},
year={2025},
month={Aug},
day={07},
volume={16},
number={1},
pages={7270},
issn={2041-1723},
doi={10.1038/s41467-025-62516-0},
url={https://doi.org/10.1038/s41467-025-62516-0}
}

@article{Zhang2025IRSSG,
  author = {Sheng Zhang and Ziyin Song and Zhong Fang and Hongming Weng and Zhijun Wang},
  title = {{IRSSG}: An open-source software package for spin space groups},
  journal = {Computer Physics Communications},
  volume = {326},
  pages = {110190},
  year = {2026},
  doi = {10.1016/j.cpc.2026.110190},
  url = {https://doi.org/10.1016/j.cpc.2026.110190}
}

@article{Gallego2016magndataI,
author = "Gallego, Samuel V. and Perez-Mato, J. Manuel and Elcoro, Luis and Tasci, Emre S. and Hanson, Robert M. and Momma, Koichi and
Aroyo, Mois I. and Madariaga, Gotzon",
title = "{{\it MAGNDATA}: towards a database of magnetic structures. I.The commensurate case}",
journal = "Journal of Applied Crystallography",
year = "2016",
volume = "49",
number = "5",
pages = "1750--1776",
month = "Oct",
doi = {10.1107/S1600576716012863},
url = {https://doi.org/10.1107/S1600576716012863},
}

@article{Gallego2016magndataII,
  title = {{MAGNDATA}: towards a database of magnetic structures. II. The incommensurate case},
  author = {Gallego, Samuel V. and Perez-Mato, J. Manuel and Elcoro, Luis and Tasci, Emre S. and Hanson, Robert M. and Aroyo, Mois I. and Madariaga, Gotzon},
  journal = {Journal of Applied Crystallography},
  volume = {49},
  number = {6},
  pages = {1941--1956},
  year = {2016},
  doi = {10.1107/S1600576716015491},
  url = {https://doi.org/10.1107/S1600576716015491}
}

@article{Jain2013MaterialsProject,
  title = {Commentary: The Materials Project: A materials genome approach to accelerating materials innovation},
  author = {Jain, Anubhav and Ong, Shyue Ping and Hautier, Geoffroy and Chen, Wei and Richards, William Davidson and Dacek, Stephen and Cholia, Shreyas and Gunter, Dan and Skinner, David and Ceder, Gerbrand and Persson, Kristin A.},
  journal = {APL Materials},
  volume = {1},
  number = {1},
  pages = {011002},
  year = {2013},
  doi = {10.1063/1.4812323},
  url = {https://doi.org/10.1063/1.4812323}
}

@article{Horton2025MaterialsProject,
  title = {Accelerated data-driven materials science with the Materials Project},
  author = {Horton, Matthew K. and Huck, Patrick and Yang, Ruo Xi and Munro, Jason M. and Dwaraknath, Shyam and Ganose, Alex M. and Kingsbury, Ryan S. and Wen, Mingjian and Shen, Jimmy X. and Mathis, Tyler S. and Kaplan, Aaron D. and Berket, Karlo and Riebesell, Janosh and George, Janine and Rosen, Andrew S. and Spotte-Smith, Evan W. C. and McDermott, Matthew J. and Cohen, Orion A. and Dunn, Alex and Kuner, Matthew C. and Rignanese, Gian-Marco and Petretto, Guido and Waroquiers, David and Griffin, Sinead M. and Neaton, Jeffrey B. and Chrzan, Daryl C. and Asta, Mark and Hautier, Geoffroy and Cholia, Shreyas and Ceder, Gerbrand and Ong, Shyue Ping and Jain, Anubhav and Persson, Kristin A.},
  journal = {Nature Materials},
  volume = {24},
  number = {10},
  pages = {1522--1532},
  year = {2025},
  doi = {10.1038/s41563-025-02272-0},
  url = {https://doi.org/10.1038/s41563-025-02272-0}
}

@misc{oddSSG,
      title={Spin Group Symmetry Criteria for Odd-parity Magnets}, 
      author={Xun-Jiang Luo and Jin-Xin Hu and Meng-Li Hu and K. T. Law},
      year={2025},
      eprint={2510.05512},
      archivePrefix={arXiv},
      primaryClass={cond-mat.other},
      url={https://arxiv.org/abs/2510.05512}, 
}

@article{SongUnifiedPRX,
  title = {Unified Symmetry Classification of Magnetic Orders via Spin Space Groups: Prediction of Coplanar Even-Wave Phases},
  author = {Song, Ziyin and Qi, Ziyue and Fang, Chen and Fang, Zhong and Weng, Hongming},
  journal = {Phys. Rev. X},
  volume = {16},
  issue = {3},
  pages = {031038},
  numpages = {12},
  year = {2026},
  month = {Aug},
  publisher = {American Physical Society},
  doi = {10.1103/zy7s-j86r},
  url = {https://link.aps.org/doi/10.1103/zy7s-j86r}
}

@misc{Luo2026Unconventional,
  title = {Unconventional Magnetism: Symmetry Classification, Hybrid-parity and Unconstrained-parity Classes},
  author = {Luo, Xun-Jiang and Li, Dan and Xiao, Rui-Chun and Shao, Ding-Fu and Li, Lei and Tian, Mingliang and Yao, Yugui},
  year = {2026},
  eprint = {2605.21336},
  archivePrefix = {arXiv},
  primaryClass = {cond-mat.mtrl-sci},
  url = {https://arxiv.org/abs/2605.21336}
}

@misc{Lange2026AntiSpinLaue,
  title = {Anti-spin Laue groups: classification of anti-altermagnets and their representative minimal models},
  author = {Lange, Colin and Jaeschke-Ubiergo, Rodrigo and Mook, Alexander and Sinova, Jairo},
  year = {2026},
  eprint = {2608.19056},
  archivePrefix = {arXiv},
  primaryClass = {cond-mat.mtrl-sci},
  url = {https://arxiv.org/abs/2608.19056}
}

@article{Li2025MagneticGroundStates,
  title = {Symmetry-guided prediction of magnetic-ordered ground states},
  author = {Li, Yuhui and Zeng, Sike and Yu, Yutong and Xiong, Renzheng and Zhao, Yu-Jun and Chen, Xiaobing and Liu, Qihang},
  journal = {Physical Review X},
  year = {2026},
  note = {Accepted 3 September 2026},
  doi = {10.1103/8ftb-swx9},
  url = {https://doi.org/10.1103/8ftb-swx9}
}

@article{Dudarev1998DFTU,
  title = {Electron-energy-loss spectra and the structural stability of nickel oxide: An LSDA+$U$ study},
  author = {Dudarev, S. L. and Botton, G. A. and Savrasov, S. Y. and Humphreys, C. J. and Sutton, A. P.},
  journal = {Phys. Rev. B},
  volume = {57},
  issue = {3},
  pages = {1505--1509},
  year = {1998},
  month = {Jan},
  publisher = {American Physical Society},
  doi = {10.1103/PhysRevB.57.1505},
  url = {https://link.aps.org/doi/10.1103/PhysRevB.57.1505}
}

@article{Kresse1996VASP,
  title = {Efficient iterative schemes for ab initio total-energy calculations using a plane-wave basis set},
  author = {Kresse, G. and Furthm{"u}ller, J.},
  journal = {Phys. Rev. B},
  volume = {54},
  issue = {16},
  pages = {11169--11186},
  year = {1996},
  month = {Oct},
  publisher = {American Physical Society},
  doi = {10.1103/PhysRevB.54.11169},
  url = {https://link.aps.org/doi/10.1103/PhysRevB.54.11169}
}

@article{Blochl1994PAW,
  title = {Projector augmented-wave method},
  author = {Bl{"o}chl, P. E.},
  journal = {Phys. Rev. B},
  volume = {50},
  issue = {24},
  pages = {17953--17979},
  year = {1994},
  month = {Dec},
  publisher = {American Physical Society},
  doi = {10.1103/PhysRevB.50.17953},
  url = {https://link.aps.org/doi/10.1103/PhysRevB.50.17953}
}

@article{Kresse1999PAW,
  title = {From ultrasoft pseudopotentials to the projector augmented-wave method},
  author = {Kresse, G. and Joubert, D.},
  journal = {Phys. Rev. B},
  volume = {59},
  issue = {3},
  pages = {1758--1775},
  year = {1999},
  month = {Jan},
  publisher = {American Physical Society},
  doi = {10.1103/PhysRevB.59.1758},
  url = {https://link.aps.org/doi/10.1103/PhysRevB.59.1758}
}

@article{Perdew1996PBE,
  title = {Generalized Gradient Approximation Made Simple},
  author = {Perdew, John P. and Burke, Kieron and Ernzerhof, Matthias},
  journal = {Phys. Rev. Lett.},
  volume = {77},
  issue = {18},
  pages = {3865--3868},
  year = {1996},
  month = {Oct},
  publisher = {American Physical Society},
  doi = {10.1103/PhysRevLett.77.3865},
  url = {https://link.aps.org/doi/10.1103/PhysRevLett.77.3865}
}

@misc{Zhou2026MagneticStructuresDatabase,
  title = {Magnetic Structures Database from Symmetry-aided High-Throughput Calculations},
  author = {Zhou, Hanjing and Mu, Yuxuan and Zhang, Dingwen and Chu, Hangbing and Kan, Erjun and Duan, Chun-Gang and Wang, Di and Liu, Huimei and Gong, Xin-Gao and Wan, Xiangang},
  year = {2026},
  eprint = {2601.01617},
  archivePrefix = {arXiv},
  primaryClass = {cond-mat.mtrl-sci},
  doi = {10.48550/arXiv.2601.01617},
  url = {https://arxiv.org/abs/2601.01617}
}

@misc{Ryu2026MixedParityTetraborides,
  title = {Unconventional Mixed-Parity Magnetism in Rare-Earth Tetraborides},
  author = {Ryu, Dong-Choon and Han, Jae-Ho and Kim, Bongjae and Kang, Chang-Jong},
  year = {2026},
  eprint = {2607.02117},
  archivePrefix = {arXiv},
  primaryClass = {cond-mat.str-el},
  doi = {10.48550/arXiv.2607.02117},
  url = {https://arxiv.org/abs/2607.02117}
}

@misc{Zhuang2026MixedParityCollinear,
  title = {Mixed-Parity Altermagnetism in Collinear Spin-Orbital Magnets},
  author = {Zhuang, Zheng-Yang and Hu, Jin-Xin and Zhang, Song-Bo and Hu, Lun-Hui and Yan, Zhongbo},
  year = {2026},
  eprint = {2605.05205},
  archivePrefix = {arXiv},
  primaryClass = {cond-mat.mes-hall},
  doi = {10.48550/arXiv.2605.05205},
  url = {https://arxiv.org/abs/2605.05205}
}

@article{Wang2025SpinOrbitalAltermagnetism,
  title = {Spin-Orbital Altermagnetism},
  author = {Wang, Zi-Ming and Zhang, Yang and Zhang, Song-Bo and Sun, Jin-Hua and Dagotto, Elbio and Xu, Dong-Hui and Hu, Lun-Hui},
  journal = {Phys. Rev. Lett.},
  volume = {135},
  number = {17},
  pages = {176705},
  year = {2025},
  doi = {10.1103/cjzw-j4v7},
  url = {https://journals.aps.org/prl/abstract/10.1103/cjzw-j4v7}
}

@article{Ma2026HyperspinAltermagnets,
  title = {Hyperspin Altermagnets},
  author = {Ma, Hai-Yang and Li, Yuanchang and Xu, Hu and Zhang, Shengbai and Jia, Jin-Feng},
  journal = {Phys. Rev. Lett.},
  volume = {137},
  number = {11},
  pages = {116704},
  year = {2026},
  doi = {10.1103/dnr2-gv31},
  url = {https://journals.aps.org/prl/abstract/10.1103/dnr2-gv31}
}

@article{Urru2025BiFeO3NodalSurfaces,
  title = {{$G$}-type antiferromagnetic {BiFeO$_3$} is a multiferroic {$g$}-wave altermagnet},
  author = {Urru, Andrea and Seleznev, Daniel and Teng, Yujia and Park, Se Young and Reyes-Lillo, Sebastian E. and Rabe, Karin M.},
  journal = {Phys. Rev. B},
  volume = {112},
  number = {10},
  pages = {104411},
  year = {2025},
  doi = {10.1103/v3fg-6smc},
  url = {https://journals.aps.org/prb/abstract/10.1103/v3fg-6smc}
}

@misc{Shen2026NiAsMixedParity,
  title = {Global magnetic phase diagram and multiple unconventional magnets in {NiAs}-type compounds},
  author = {Shen, Shibo and Wang, Yilin},
  year = {2026},
  eprint = {2605.28391},
  archivePrefix = {arXiv},
  primaryClass = {cond-mat.mtrl-sci},
  doi = {10.48550/arXiv.2605.28391},
  url = {https://arxiv.org/abs/2605.28391}
}

@article{Zhu2025MagneticGeometry,
  title = {Magnetic geometry induced quantum geometry and nonlinear transports},
  author = {Zhu, Haiyuan and Li, Jiayu and Chen, Xiaobing and Yu, Yutong and Liu, Qihang},
  journal = {Nature Communications},
  volume = {16},
  pages = {4882},
  year = {2025},
  doi = {10.1038/s41467-025-60128-2},
  url = {https://www.nature.com/articles/s41467-025-60128-2}
}

@misc{Hu2026NiS2,
  title = {Non-collinear Altermagnetic Phases in the {Mott} Insulator {NiS$_2$}},
  author = {Hu, Mengli and Iraola, Mikel I. and McClarty, Paul and van den Brink, Jeroen and Vergniory, Maia G.},
  year = {2026},
  eprint = {2603.01329},
  archivePrefix = {arXiv},
  primaryClass = {cond-mat.mtrl-sci},
  doi = {10.48550/arXiv.2603.01329},
  url = {https://arxiv.org/abs/2603.01329}
}

\clearpage
% ==================== SUPPLEMENTARY INFORMATION ====================
\setcounter{section}{0}
\setcounter{subsection}{0}
\setcounter{subsubsection}{0}
\setcounter{equation}{0}
\setcounter{figure}{0}
\setcounter{table}{0}
\renewcommand{\theequation}{S\arabic{equation}}
\renewcommand{\thefigure}{S\arabic{figure}}
\renewcommand{\thetable}{S\arabic{table}}
\renewcommand{\theHequation}{supp.\arabic{equation}}
\renewcommand{\theHfigure}{supp.\arabic{figure}}
\renewcommand{\theHtable}{supp.\arabic{table}}
\renewcommand{\theHsection}{supp.\arabic{section}}
\setcounter{tocdepth}{2}
\let\standardSubsection\subsection
\renewcommand{\subsection}[1]{\standardSubsection{#1}\leavevmode\par}
\let\standardSubsubsection\subsubsection
\renewcommand{\subsubsection}[1]{\standardSubsubsection{#1}\leavevmode\par}
\section*{Supplementary Information}
\begin{center}
\large Crystal symmetry predicts unconventional magnetism
\end{center}
\renewcommand{\contentsname}{Supplementary contents}
\tableofcontents

\clearpage

%################################################################################
%##### SECTION: Overview and prediction problem
%################################################################################

\section{Overview and prediction problem}

Unconventional magnetic states have attracted rapidly growing interest because they extend magnetic functionality beyond the conventional distinction between ferromagnets and antiferromagnets. Representative examples include altermagnets, $p$-wave magnets and more general compensated magnetic states with non-trivial momentum-dependent spin textures. Although these systems can have vanishing net magnetization, their magnetic symmetry permits pronounced spin polarization in momentum space and can generate characteristic electronic and transport phenomena, including non-relativistic spin splitting, spin-polarized transport, anomalous and nonlinear Hall responses, and topological electronic or magnonic states. These developments have considerably broadened the theoretical landscape of magnetic materials. Experimentally, however, the number of materials for which unconventional magnetic order has been firmly established remains comparatively limited.

A major bottleneck is the determination of the magnetic structure itself. Whether a material belongs to a conventional or unconventional magnetic class depends on how its magnetic moments are arranged in real space and on the symmetry of the resulting magnetic state. Complete magnetic structures are substantially more difficult to establish than non-magnetic crystal structures. Their experimental determination often requires magnetic diffraction or other dedicated probes, sufficiently large and high-quality samples, and careful refinement of the magnetic propagation vector, magnetic unit cell and moment directions. Consequently, experimentally refined magnetic structures are available for only a small subset of known magnetic compounds. By contrast, crystallographic structures without magnetic-moment information are routinely measured and are available on a much larger scale in experimental and computational materials databases.

The same difficulty appears from the theoretical side. Predicting the magnetic ground state of a material is generally a substantially more demanding problem than determining its non-magnetic crystal structure. A realistic material can support many competing magnetic configurations that differ in propagation vector, magnetic-cell size, moment direction, collinearity and local-moment amplitude. Their relative energies may depend sensitively on the exchange--correlation functional, Hubbard interaction parameters, spin--orbit coupling, structural relaxation, magnetic initialization and other details of the calculation. In systems with several nearly competing states, different computational settings can therefore lead to different energetic orderings.

These considerations motivate a different prediction problem. Rather than first determining the unique magnetic ground state and only then asking whether it is unconventional, we ask whether the unconventional magnetic character can already be inferred from the symmetry-compatible magnetic possibilities allowed by the crystallographic structure. The key observation is that the detailed real-space magnetic configuration and the reciprocal-space spin texture do not contain identical information. Different real-space magnetic configurations can belong to different spin-space groups (SSGs) while nevertheless sharing the same broad type of momentum-space spin polarization. In particular, symmetry can constrain whether a magnetic state permits a conventional zero-dimensional spin texture or an unconventional finite-dimensional texture, and in the one-dimensional case can further determine the symmetry character of the spin polarization through $p$-, $d$-, $f$-, $g$- or $i$-wave basis functions.

This reciprocal-space characterization is especially relevant because it is closely connected to experimentally observable electronic properties. Momentum-dependent spin polarization determines the symmetry of non-relativistic spin splitting and strongly constrains spin-dependent electronic structure and transport. It therefore provides a natural level at which to characterize unconventional magnetism even when the precise microscopic arrangement of moments has not yet been resolved. Importantly, the symmetry character of the spin texture can be shared by several competing magnetic configurations. In such cases, the detailed magnetic ground state may remain sensitive to microscopic energetics, whereas the unconventional magnetic character is already common to the entire symmetry-compatible candidate ensemble.

The strategy of this work is therefore to predict magnetic character at the level of a candidate ensemble rather than to assume a single magnetic configuration. Starting from a non-magnetic crystallographic structure and a specified magnetic sublattice, we enumerate SSGs within a controlled search space and explicitly construct the corresponding zero-net-moment magnetic structures. In the present implementation, we consider one magnetic chemical species occupying a single symmetry-equivalent crystallographic orbit, commensurate magnetic structures with magnetic-cell index $I_k\leq4$, and candidate SSGs satisfying the spatial-symmetry condition imposed by the parent crystallographic structure. Each retained candidate is verified to have vanishing total moment and the intended full SSG symmetry.

We then classify every candidate using both real-space and reciprocal-space information. The real-space configuration is labelled as collinear, coplanar or noncoplanar, while the reciprocal-space spin texture is classified by its symmetry-allowed dimensionality. Zero-dimensional textures are identified as conventional compensated magnetic states, whereas finite-dimensional textures are classified as unconventional. One-dimensional textures form the spin-polarized wave-like subset and are further characterized by their symmetry basis functions. These candidate-level classifications are subsequently aggregated at the material level to quantify how strongly unconventional magnetic character is constrained across the symmetry-compatible ensemble.

To test whether this ensemble-level information is experimentally meaningful, we benchmark the framework against magnetic structures collected from MAGNDATA and compare the symmetry-generated candidate ensembles with experimentally established magnetic states. We then apply the same procedure on a much larger scale to magnetic materials collected from the Materials Project, where detailed experimental magnetic structures are generally unavailable, and identify a large set of materials with strongly constrained unconventional candidate ensembles. The complete benchmark construction, high-throughput statistics and material classifications are presented in the following sections.

%################################################################################
%##### SECTION: Spin-space groups and their action on magnetic structures
%################################################################################

\section{Spin-space groups and their action on magnetic structures}

\subsection{Spin-space-group symmetry}

Spin-space groups (SSGs) provide a symmetry framework for describing magnetic structures in which spin and real-space transformations are treated as independent operations. For an ordinary space group, a symmetry operation acts only on the spatial coordinates of the crystal. In a conventional magnetic space group (MSG), the transformation of an axial magnetic moment is tied to the corresponding spatial operation. In the absence of spin--orbit coupling, however, there is no fundamental requirement that a rotation in real space be accompanied by the same rotation in spin space. The spin and lattice degrees of freedom can therefore transform independently, and the resulting enlarged symmetry is described by an SSG \cite{jiang2023enumeration,SongUnifiedPRX}.

A general SSG operation is written as
\begin{equation}
    g=\{U\Vert R|\boldsymbol{\tau}\},
    \label{eq:ssg_operation}
\end{equation}
where $U\in O(3)$ acts in spin space, while $\{R|\boldsymbol{\tau}\}$ is a spatial operation consisting of a point operation $R\in O(3)$ and a translation $\boldsymbol{\tau}$. Throughout this work, we use the double vertical line $\Vert$ to separate the spin-space operation from the real-space operation. This notation makes explicit that $U$ and $R$ are, in general, independent.

For $\det(U)=+1$, $U$ is a proper spin rotation and the corresponding operation is unitary. An improper spin-space operation with $\det(U)=-1$ is understood to contain time reversal. Equivalently, it may be written as
\begin{equation}
    U=\widetilde{U}\mathcal{T},
    \qquad
    \widetilde{U}=\det(U)U\in SO(3),
    \label{eq:improper_spin_operation}
\end{equation}
where $\mathcal{T}$ denotes time reversal. This convention allows both unitary and antiunitary spin transformations to be represented uniformly by matrices in $O(3)$.

An SSG can therefore be written schematically as
\begin{equation}
    \mathcal{G}^{S}
    =
    \bigcup_i
    \{U_i\Vert R_i|\boldsymbol{\tau}_i\}\mathcal{T}_{H},
    \label{eq:ssg_group}
\end{equation}
where $\mathcal{T}_{H}$ denotes the translation subgroup associated with the magnetic structure. The collection of spatial parts $\{R_i|\boldsymbol{\tau}_i\}$ forms a space group $G$. Throughout this work, we refer to this spatial projection of the SSG as the \emph{parent space group} (parent SG). The parent SG specifies the complete set of spatial operations that appear in the SSG after the spin-space parts are removed. This definition is particularly important for the prediction scheme developed below, in which candidate magnetic SSGs are selected by requiring their parent SG to coincide with the space group of the corresponding non-magnetic crystal structure. Unless otherwise stated, the term parent SG will be used throughout the Supplementary Information to denote this spatial projection $G$ of an SSG.

Within the parent SG $G$, the operations with a trivial spin part,
\begin{equation}
    H=
    \left\{
    \{E\Vert R|\boldsymbol{\tau}\}
    \in \mathcal{G}^{S}
    \right\},
    \label{eq:pure_lattice_group}
\end{equation}
form an invariant subgroup $H\triangleleft G$. The quotient group
\begin{equation}
    Q=G/H
    \label{eq:ssg_quotient}
\end{equation}
describes how the spatial cosets are associated with nontrivial transformations in spin space. This quotient-group construction provides the basis for the systematic enumeration of SSGs.

Another important subgroup is the spin-only group,
\begin{equation}
    S_0=
    \left\{
    U\; \middle|\;
    \{U\Vert E|\mathbf{0}\}\in\mathcal{G}^{S}
    \right\},
    \label{eq:spin_only_group}
\end{equation}
whose operations act on the magnetic moments without moving the atomic positions. The form of $S_0$ is closely connected to the geometry of the magnetic configuration. For example, collinear and coplanar magnetic structures possess characteristic nontrivial spin-only symmetries, whereas a generic noncoplanar magnetic structure has no nontrivial continuous spin-only symmetry. These constraints will be used in the next section to classify magnetic structures in real and reciprocal spaces.

The SSG description is particularly natural in the weak-spin--orbit-coupling limit. In this limit, the electronic Hamiltonian can be viewed schematically as
\begin{equation}
    \hat{H}
    =
    \frac{\hat{\mathbf p}^{\,2}}{2m}
    +
    V(\mathbf r)
    +
    \mathbf M(\mathbf r)\cdot\hat{\mathbf s},
    \label{eq:nonrelativistic_magnetic_hamiltonian}
\end{equation}
where $V(\mathbf r)$ is the non-magnetic crystal potential and $\mathbf M(\mathbf r)$ represents the ordered magnetic background. Because the spin--orbit interaction that locks spin rotations to lattice rotations is absent in Eq.~\eqref{eq:nonrelativistic_magnetic_hamiltonian}, the Hamiltonian can possess SSG symmetries that are absent from its relativistic MSG description. An SSG always characterizes the symmetry of a magnetic configuration itself; for electronic structures, it should be regarded as an exact symmetry in the non-relativistic limit and as an approximate symmetry when spin--orbit coupling is weak.

\subsection{SSG notation and numerical labels}

The systematic enumeration of SSGs requires a notation that distinguishes groups generated from the same parent SG but with different magnetic translation subgroups and different spin-space representations. We follow the numerical convention introduced in Ref.~\cite{jiang2023enumeration}. In the notation used throughout this work, an SSG is labelled as
\begin{equation}
    N_{\mathrm{SG}}.I_k.I_t.N_{\mathrm{rep}},
    \label{eq:ssg_number_general}
\end{equation}
with an additional suffix ``L'' or ``P'' for collinear and coplanar SSGs, respectively.

The first number, $N_{\mathrm{SG}}$, is the International Tables number of the parent SG $G$, namely the spatial projection of the SSG. Thus, all SSGs beginning with the same $N_{\mathrm{SG}}$ share the same parent SG, although their spin-space operations, magnetic translation subgroups and magnetic configurations can differ.

To define the next two indices, we denote the pure-lattice invariant subgroup by $H\triangleleft G$. Let $T_G$ and $T_H$ be the translation subgroups of $G$ and $H$, respectively, and let $P_G$ and $P_H$ denote their corresponding point groups. The two indices are
\begin{equation}
    I_k=|T_G/T_H|,
    \qquad
    I_t=|P_G/P_H|.
    \label{eq:ik_it_definition}
\end{equation}
Here $I_k$ is the $k$ index, which measures the enlargement of the magnetic translation cell, whereas $I_t$ is the $t$ index associated with the reduction of the point-group part of the pure-lattice subgroup. Together, $I_k$ and $I_t$ determine the index of $H$ in the parent SG $G$,
\begin{equation}
    |G/H|=I_k I_t.
    \label{eq:quotient_index}
\end{equation}

For a fixed $N_{\mathrm{SG}}$, $I_k$ and $I_t$, more than one inequivalent real representation of the quotient group $Q=G/H$ may generate an SSG. The fourth number, $N_{\mathrm{rep}}$, enumerates these inequivalent representations. For a generic noncoplanar SSG it labels a three-dimensional real representation. For collinear and coplanar SSGs, the corresponding one- and two-dimensional representation constructions are distinguished by the suffixes
\begin{equation}
    N_{\mathrm{SG}}.I_k.I_t.N_{\mathrm{rep}}.\mathrm{L}
    \label{eq:ssg_number_collinear}
\end{equation}
and
\begin{equation}
    N_{\mathrm{SG}}.I_k.I_t.N_{\mathrm{rep}}.\mathrm{P},
    \label{eq:ssg_number_coplanar}
\end{equation}
respectively. No additional suffix is used for a generic noncoplanar SSG.

As an example, the SSG
\begin{equation}
    223.1.2.3.\mathrm{L}
\end{equation}
has parent SG~223, magnetic-cell index $I_k=1$, $t$ index $I_t=2$, and corresponds to the third relevant one-dimensional representation for this pair of indices; the suffix L indicates a collinear SSG. Similarly,
\begin{equation}
    223.2.4.1.\mathrm{P}
\end{equation}
has parent SG~223 and $I_k=2$, and describes a coplanar SSG. This numerical notation is used throughout the high-throughput calculations because it provides a unique identifier that can be directly connected to the enumerated SSG database.

The international symbols in Table~\ref{tab:ssg_international_symbols} encode paired generators: $x^{\delta}$ denotes $\{\delta\Vert x\}$, combining the spatial generator $x$ with the spin operation $\delta$, while the three superscripts on the lattice symbol specify the spin operations paired with the three standard translation generators\cite{Zhang2025IRSSG}.

\begin{table}[p]
    \centering
    \small\setstretch{1.1}
    \setlength{\tabcolsep}{4pt}
    \caption{Numerical labels and IRSSG international symbols\cite{Zhang2025IRSSG} of the symmetry-generated candidates for \ch{VGe3} and tetragonal \ch{Fe2SiO4}. Candidate indices follow Tables~\ref{tab:vge3_candidates} and~\ref{tab:fe2sio4_screening}. Symbols are given in the database standard setting.}
    \label{tab:ssg_international_symbols}
    \begin{tabular}{@{}lllc@{}}
        \toprule
        Material & Candidate & Numerical SSG label & International symbol \\
        \midrule
        \ch{VGe3} & 01 & 223.1.2.3.L & $\mathrm{P}^{1,1,1}\,\mathrm{m}^{1}\,\bar{3}^{1}\,\mathrm{n}^{\bar{1}}\,\left(C_{i}^{\mathrm{I}}\right)$ \\
         & 02 & 223.2.4.1.P & $\mathrm{P}^{2,2,2}\,\mathrm{m}^{m}\,\bar{3}^{m}\,\mathrm{n}^{4}\,\left(C_{4v}^{\mathrm{II}}\right)$ \\
         & 03 & 223.2.4.4 & $\mathrm{P}^{2,2,2}\,\mathrm{m}^{m}\,\bar{3}^{m}\,\mathrm{n}^{\bar{4}^{-1}}\,\left(D_{2d}^{\mathrm{III}}\right)$ \\
        \midrule
        \ch{Fe2SiO4} & 01 & 141.1.2.6.L & $\mathrm{I}^{1,1,1}\,4_1^{\bar{1}}/\,\mathrm{a}^{1}\,\mathrm{m}^{1}\,\mathrm{d}^{\bar{1}}\,\left(C_{i}^{\mathrm{I}}\right)$ \\
         & 02 & 141.1.8.2.P & $\mathrm{I}^{1,1,1}\,4_1^{4}/\,\mathrm{a}^{2}\,\mathrm{m}^{m}\,\mathrm{d}^{m}\,\left(C_{4v}^{\mathrm{II}}\right)$ \\
         & 03 & 141.1.8.34 & $\mathrm{I}^{1,1,1}\,4_1^{\bar{4}^{-1}}/\,\mathrm{a}^{2}\,\mathrm{m}^{m}\,\mathrm{d}^{2}\,\left(D_{2d}^{\mathrm{III}}\right)$ \\
         & 04 & 141.2.4.1.P & $\mathrm{I}^{2,2,2}\,4_1^{4^{-1}}/\,\mathrm{a}^{m}\,\mathrm{m}^{1}\,\mathrm{d}^{4}\,\left(C_{4v}^{\mathrm{II}}\right)$ \\
         & 05 & 141.2.4.3.P & $\mathrm{I}^{2,2,2}\,4_1^{m}/\,\mathrm{a}^{m}\,\mathrm{m}^{m}\,\mathrm{d}^{4^{-1}}\,\left(C_{4v}^{\mathrm{II}}\right)$ \\
         & 06 & 141.2.4.12 & $\mathrm{I}^{2,2,2}\,4_1^{2}/\,\mathrm{a}^{m}\,\mathrm{m}^{m}\,\mathrm{d}^{\bar{4}}\,\left(D_{2d}^{\mathrm{III}}\right)$ \\
         & 07 & 141.2.4.2 & $\mathrm{I}^{2,2,2}\,4_1^{\bar{4}}/\,\mathrm{a}^{m}\,\mathrm{m}^{1}\,\mathrm{d}^{\bar{4}^{-1}}\,\left(D_{2d}^{\mathrm{III}}\right)$ \\
         & 08 & 141.3.2.1.P & $\mathrm{I}^{3,3,3^{-1}}\,4_1^{3^{-1}}/\,\mathrm{a}^{m}\,\mathrm{m}^{1}\,\mathrm{d}^{1}\,\left(C_{3v}^{\mathrm{II}}\right)$ \\
         & 09 & 141.3.4.2.P & $\mathrm{I}^{3,3,3^{-1}}\,4_1^{6}/\,\mathrm{a}^{m}\,\mathrm{m}^{1}\,\mathrm{d}^{2}\,\left(C_{6v}^{\mathrm{II}}\right)$ \\
         & 10 & 141.3.4.11 & $\mathrm{I}^{3,3,3^{-1}}\,4_1^{\bar{3}^{-1}}/\,\mathrm{a}^{m}\,\mathrm{m}^{1}\,\mathrm{d}^{\bar{1}}\,\left(D_{3d}^{\mathrm{III}}\right)$ \\
         & 11 & 141.3.4.7 & $\mathrm{I}^{3,3,3^{-1}}\,4_1^{\bar{6}}/\,\mathrm{a}^{m}\,\mathrm{m}^{1}\,\mathrm{d}^{m}\,\left(D_{3h}^{\mathrm{III}}\right)$ \\
         & 12 & 141.4.4.1.P & $\mathrm{I}^{4^{-1},4^{-1},4}\,4_1^{8^{-1}}/\,\mathrm{a}^{m}\,\mathrm{m}^{1}\,\mathrm{d}^{8^{-3}}\,\left(C_{8v}^{\mathrm{II}}\right)$ \\
         & 13 & 141.4.4.2.P & $\mathrm{I}^{4,4,4^{-1}}\,4_1^{8^{-3}}/\,\mathrm{a}^{m}\,\mathrm{m}^{1}\,\mathrm{d}^{8^{-1}}\,\left(C_{8v}^{\mathrm{II}}\right)$ \\
         & 14 & 141.4.4.3 & $\mathrm{I}^{4^{-1},4^{-1},4}\,4_1^{\bar{8}^{3}}/\,\mathrm{a}^{m}\,\mathrm{m}^{1}\,\mathrm{d}^{\bar{8}}\,\left(D_{4d}^{\mathrm{III}}\right)$ \\
         & 15 & 141.4.4.7 & $\mathrm{I}^{4,4,4^{-1}}\,4_1^{\bar{8}}/\,\mathrm{a}^{m}\,\mathrm{m}^{1}\,\mathrm{d}^{\bar{8}^{3}}\,\left(D_{4d}^{\mathrm{III}}\right)$ \\
        \bottomrule
    \end{tabular}
\end{table}

\subsection{Magnetic-cell index \texorpdfstring{$\Ik$}{Ik}}

Among the indices entering the SSG label, $I_k$ plays a particularly important role in the present work. As defined in Eq.~\eqref{eq:ik_it_definition},
\begin{equation}
    I_k=|T_G/T_H|,
    \label{eq:ik_definition_again}
\end{equation}
where $T_G$ is the translation group of the parent SG and $T_H$ is the subgroup of translations that appear as pure-lattice operations in the SSG.

Physically, $I_k$ measures the volume enlargement of the magnetic unit cell relative to the crystallographic primitive cell. For
\begin{equation}
    I_k=1,
\end{equation}
the magnetic order does not enlarge the translational unit cell. The magnetic propagation is therefore compatible with the crystallographic translation lattice. For $I_k>1$, some crystallographic translations are accompanied by nontrivial spin-space operations and are no longer pure translations of the magnetic structure. The magnetic unit cell is consequently enlarged, with its volume being $I_k$ times that of the crystallographic primitive cell.

For example, an antiferromagnetic structure in which a crystallographic translation exchanges two oppositely oriented magnetic sublattices may have $I_k=2$. The corresponding translation remains part of the parent SG $G$, but it is accompanied by a nontrivial spin operation in the full SSG and therefore does not belong to $T_H$.

The distinction between $I_k=1$ and $I_k>1$ is especially useful for the prediction problem considered here. Determining the complete orientation of all magnetic moments generally requires a full magnetic-structure refinement. By contrast, whether a magnetic order enlarges the crystallographic unit cell can often be established more readily from magnetic diffraction through the presence or absence of additional magnetic Bragg vectors. Knowledge of $I_k$, even without knowledge of the complete magnetic configuration, can therefore substantially reduce the symmetry-compatible candidate space.

The complete SSG enumeration contains magnetic structures with substantially larger values of $I_k$. In the present high-throughput search, however, we restrict the candidate set to
\begin{equation}
    I_k\leq4.
    \label{eq:ik_cutoff}
\end{equation}
This cutoff covers nearly all experimentally reported magnetic structures in the MAGNDATA dataset considered in this work. Among the 1,626 MAGNDATA entries used to construct the benchmark dataset, only 10 have experimentally reported SSGs with $I_k>4$. Moreover, the overwhelming majority involve only very small magnetic-cell enlargements: 1,570 of the 1,626 entries have $I_k\leq2$. The $I_k\leq4$ restriction therefore captures essentially the experimentally relevant range of commensurate magnetic-cell sizes while substantially reducing the number of symmetry candidates that need to be generated in the high-throughput search. Magnetic structures with larger $I_k$ remain outside the present search space and are discussed further in the section on scope and limitations.

\subsection{Action of SSG operations on real-space magnetic structures}

A magnetic structure can be represented by a periodic spin or magnetization field $\mathbf S(\mathbf r)$. For an SSG operation
\begin{equation}
    g=\{U\Vert R|\boldsymbol{\tau}\},
\end{equation}
we define its spatial part as
\begin{equation}
    g_r=\{R|\boldsymbol{\tau}\}.
\end{equation}
The action of the SSG operation on the real-space spin configuration is
\begin{equation}
    \{U\Vert R|\boldsymbol{\tau}\}\mathbf S(\mathbf r)
    =
    U\,\mathbf S(g_r^{-1}\mathbf r).
    \label{eq:ssg_real_space_action}
\end{equation}
A symmetry operation of the magnetic structure satisfies
\begin{equation}
    \mathbf S(\mathbf r)
    =
    U\,\mathbf S(g_r^{-1}\mathbf r).
    \label{eq:ssg_real_space_invariance}
\end{equation}

Equation~\eqref{eq:ssg_real_space_invariance} has a direct interpretation for localized magnetic moments. Suppose that the spatial operation maps a site $\mathbf r_i$ onto another site $\mathbf r_j$ up to a lattice translation,
\begin{equation}
    \mathbf r_j
    =
    g_r\mathbf r_i .
\end{equation}
Their magnetic moments must then satisfy
\begin{equation}
    \mathbf m_j
    =
    U\mathbf m_i.
    \label{eq:ssg_moment_mapping}
\end{equation}
Thus, once the magnetic moment on one reference site is specified, the SSG operations can generate the moments on all sites belonging to the same magnetic orbit.

Of particular importance are on-site operations. If the spatial part of an SSG operation leaves a site invariant modulo a lattice translation,
\begin{equation}
    g_r\mathbf r_i
    =
    \mathbf r_i+\mathbf R_n,
\end{equation}
then Eq.~\eqref{eq:ssg_real_space_invariance} reduces locally to
\begin{equation}
    \mathbf m_i=U\mathbf m_i.
    \label{eq:onsite_moment_constraint}
\end{equation}
The magnetic moment must therefore lie in the invariant subspace of the corresponding spin operation. The intersection of the invariant subspaces of all on-site operations determines the magnetic degrees of freedom allowed on that site.

This provides a particularly useful way to understand real-space magnetic geometry. A continuous spin-only rotation around a fixed axis constrains all magnetic moments to that axis and therefore produces a collinear configuration. A spin-only mirror can constrain the moments to a plane, giving a coplanar configuration. In the absence of nontrivial spin-only constraints, a generic magnetic structure may be noncoplanar. Conversely, the presence of pure time reversal as an on-site symmetry forces
\begin{equation}
    \mathbf m_i=0,
\end{equation}
and therefore forbids magnetic order on that site.

In addition to these local constraints, the complete set of spin parts $\{U\}$ appearing in the SSG constrains the possible uniform magnetization. If their common invariant subspace contains a nonzero vector, a net magnetization is symmetry allowed. If no nonzero vector remains invariant under the full spin-part point group, the magnetic order is symmetry compensated. This distinction will be used below to restrict the high-throughput construction to zero-net-moment magnetic candidates.

\subsection{Action of SSG operations on reciprocal-space spin textures}

The same SSG also constrains the momentum-dependent spin polarization of the electronic structure. We denote the spin texture in reciprocal space by $\mathbf S(\mathbf k)$. Importantly, an SSG operation does not act on $\mathbf S(\mathbf k)$ in exactly the same way as it acts on the real-space field $\mathbf S(\mathbf r)$. For
\begin{equation}
    g=\{U\Vert R|\boldsymbol{\tau}\},
\end{equation}
the transformation law is
\begin{equation}
    \{U\Vert R|\boldsymbol{\tau}\}\mathbf S(\mathbf k)
    =
    U\,
    \mathbf S\!\left[
        \det(U)R^{-1}\mathbf k
    \right].
    \label{eq:ssg_reciprocal_space_action}
\end{equation}
Equivalently, the symmetry constraint relates the spin polarizations at symmetry-related momenta as
\begin{equation}
    \mathbf S\!\left[\det(U)R\mathbf k\right]
    =U\,\mathbf S(\mathbf k).
    \label{eq:ssg_reciprocal_space_invariance}
\end{equation}
Here $R$ is expressed in Cartesian coordinates. The translation $\boldsymbol{\tau}$ does not shift momentum in this relation for the spin expectation value.

The factor $\det(U)$ is the key difference between Eqs.~\eqref{eq:ssg_real_space_action} and \eqref{eq:ssg_reciprocal_space_action}. For a unitary operation with $\det(U)=+1$, the momentum transforms according to the ordinary spatial operation. For an antiunitary operation with $\det(U)=-1$, time reversal contributes an additional reversal of momentum,
\begin{equation}
    \mathbf k\rightarrow-\mathbf k.
\end{equation}
Consequently, the same SSG operation can impose qualitatively different constraints on the spin configuration in real and reciprocal spaces.

This difference is central to unconventional magnetism. Two magnetic structures that are similar from the viewpoint of their real-space spin geometry may exhibit fundamentally different momentum-space spin textures. Conversely, distinct real-space magnetic configurations can share the same dimensionality and wave character of their reciprocal-space spin polarization. Altermagnetism, coplanar odd-wave magnetism and coplanar even-wave magnetism are examples of phases whose distinction is naturally expressed through the different SSG constraints on $\mathbf S(\mathbf r)$ and $\mathbf S(\mathbf k)$.

At a generic momentum, the dimensionality of the allowed $\mathbf S(\mathbf k)$ is determined by the SSG operations that leave that momentum invariant after the transformation in Eq.~\eqref{eq:ssg_reciprocal_space_invariance}. For such operations, the spin polarization must satisfy
\begin{equation}
    \mathbf S(\mathbf k)=U\mathbf S(\mathbf k).
    \label{eq:k_onsite_constraint}
\end{equation}
The common invariant subspace of these spin operations can have dimension zero, one, two or three. This provides the basis for the reciprocal-space classification used in this work. In particular, a one-dimensional invariant subspace constrains the spin polarization at generic momentum to a fixed spin axis, while the remaining SSG operations determine how the sign and magnitude of this polarization transform throughout the Brillouin zone. The resulting conventional, unconventional and spin-polarized magnetic classes, together with their wave-symmetry classification, are introduced in the following section.

%################################################################################
%##### SECTION: Classification of magnetic orders from SSG symmetry
%################################################################################

\section{Classification of magnetic orders from SSG symmetry}
\label{sec:ssg_classification}

The different actions of an SSG on $\mathbf S(\mathbf r)$ and $\mathbf S(\mathbf k)$ provide a natural framework for classifying magnetic orders independently in real and reciprocal spaces. In real space, the spin-only subgroup determines whether the magnetic configuration is collinear, coplanar or noncoplanar, while the spin-space symmetry determines whether a nonzero net magnetization is symmetry allowed. In reciprocal space, the SSG constrains the dimensionality and symmetry of the momentum-dependent spin polarization. Combining these complementary constraints yields a systematic classification that includes conventional antiferromagnets, altermagnets, coplanar odd- and even-wave magnets, and more general unconventional magnetic states \cite{SongUnifiedPRX}.

In the present work, we use the reciprocal-space classification as the primary criterion for distinguishing conventional and unconventional compensated magnetism. We first introduce the real-space magnetic geometry and the dimensionality of the reciprocal-space spin texture, and then define the conventional, unconventional and spin-polarized classes used throughout the high-throughput search. We further characterize the momentum dependence through odd-, even- and mixed-wave symmetries for one-dimensional textures and through the wave symmetries of individual spin components for multidimensional textures.

\subsection{Real-space classification}

As discussed in the previous section, the spin-only subgroup $S_0$ directly constrains the allowed magnetic degrees of freedom without moving the atomic positions. According to the dimension of the spin subspace allowed by $S_0$, a magnetic structure can be classified as nonmagnetic, collinear, coplanar or noncoplanar. In the magnetic cases considered here, the moments are respectively restricted to a common axis, a common plane or the full three-dimensional spin space. This real-space classification will be used throughout the following symmetry analysis.

Independently of this geometrical classification, the SSG also determines whether a nonzero net magnetization is symmetry allowed. We denote by
\begin{equation}
    P_s=
    \left\{
    U\;\middle|\;
    \{U\Vert R|\boldsymbol{\tau}\}\in\mathcal G^S
    \right\}
    \label{eq:spin_part_point_group}
\end{equation}
the point group formed by the spin-space parts of the SSG operations. A nonzero net magnetization $\mathbf M$ is symmetry allowed only if there exists a nonzero vector satisfying
\begin{equation}
    U\mathbf M=\mathbf M,
    \qquad
    \forall\,U\in P_s.
    \label{eq:uniform_magnetization_constraint}
\end{equation}
If such a common invariant vector exists, $P_s$ is polar and a nonzero net magnetization is symmetry allowed. If the only common invariant vector is the zero vector, $P_s$ is nonpolar and the SSG forbids a net magnetization. In the present work, we retain the latter class because our objective is to construct zero-net-moment magnetic candidates.

Importantly, the labels collinear, coplanar and noncoplanar describe only the geometry of the real-space magnetic configuration. They do not by themselves determine whether the corresponding electronic spin texture is conventional or unconventional. That distinction follows from the SSG constraints in reciprocal space.

\subsection{Dimensionality of reciprocal-space spin textures}

We characterize the reciprocal-space magnetic structure through the momentum-dependent spin polarization $\mathbf S(\mathbf k)$. At a generic momentum $\mathbf k$, the allowed spin polarization is constrained by the SSG operations that act locally at the same momentum. As discussed in the previous section, these include unitary operations of the form
\begin{equation}
    \{U\Vert E|\boldsymbol{\tau}\},
    \qquad
    \det(U)=+1,
    \label{eq:k_local_unitary}
\end{equation}
as well as antiunitary operations for which the momentum reversal associated with time reversal is compensated by a spatial operation, for example
\begin{equation}
    \{U\Vert P|\boldsymbol{\tau}\},
    \qquad
    \det(U)=-1,
    \label{eq:k_local_antiunitary}
\end{equation}
where $P$ denotes spatial inversion. For every such operation, the spin polarization at a generic $\mathbf k$ must satisfy
\begin{equation}
    \mathbf S(\mathbf k)=U\mathbf S(\mathbf k).
    \label{eq:k_local_spin_constraint}
\end{equation}

We define the symmetry-allowed spin subspace at generic momentum as
\begin{equation}
    V_{\mathbf k}
    =
    \bigcap_{U\in\mathcal U_{\mathbf k}}
    \mathrm{Fix}(U),
    \label{eq:allowed_spin_subspace}
\end{equation}
where $\mathcal U_{\mathbf k}$ is the set of spin-space parts of all SSG operations acting locally at generic $\mathbf k$, and
\begin{equation}
    \mathrm{Fix}(U)
    =
    \left\{
    \mathbf v\in\mathbb R^3
    \mid
    U\mathbf v=\mathbf v
    \right\}.
\end{equation}
The dimensionality of the reciprocal-space spin texture is then defined as
\begin{equation}
    d_{\mathbf k}
    =
    \dim V_{\mathbf k},
    \qquad
    d_{\mathbf k}=0,1,2,3.
    \label{eq:k_texture_dimension}
\end{equation}
Throughout this work, this quantity is denoted computationally as \texttt{k\_texture\_dim}. Because the classification is performed at a generic momentum, we suppress the subscript $\mathbf k$ below and simply denote the dimension by $d$.

For $d=0$, the symmetry constraints eliminate all nonzero spin polarization at generic momentum,
\begin{equation}
    \mathbf S(\mathbf k)=0.
    \label{eq:0d_texture}
\end{equation}
This is the reciprocal-space structure that we identify as conventional magnetism in the present work.

For $d=1$, the spin polarization is constrained to a single fixed axis $\hat{\mathbf n}$,
\begin{equation}
    \mathbf S(\mathbf k)
    =
    s(\mathbf k)\hat{\mathbf n}.
    \label{eq:1d_texture}
\end{equation}
The magnitude and sign of the scalar polarization $s(\mathbf k)$ can vary throughout the Brillouin zone, but the polarization direction is symmetry restricted to the same one-dimensional spin subspace. We refer to these states as spin-polarized magnetic states.

For $d=2$, the spin polarization is restricted to a fixed plane but is not confined to a single axis. Choosing the plane as the $xy$ plane,
\begin{equation}
    \mathbf S(\mathbf k)
    =
    S_x(\mathbf k)\hat{\mathbf x}
    +
    S_y(\mathbf k)\hat{\mathbf y}.
    \label{eq:2d_texture}
\end{equation}
The spin orientation may therefore rotate within this plane as a function of momentum. We refer to this as a two-dimensional or spin-coplanar reciprocal-space texture.

Finally, for $d=3$, no SSG symmetry restricts the generic spin polarization to a lower-dimensional spin subspace. In general,
\begin{equation}
    \mathbf S(\mathbf k)
    =
    S_x(\mathbf k)\hat{\mathbf x}
    +
    S_y(\mathbf k)\hat{\mathbf y}
    +
    S_z(\mathbf k)\hat{\mathbf z},
    \label{eq:3d_texture}
\end{equation}
and the spin can point along an arbitrary direction at generic momentum. This is a three-dimensional or spin-noncoplanar reciprocal-space texture.

The real-space and reciprocal-space dimensions need not coincide. A collinear real-space magnetic order may have either a vanishing or a one-dimensional reciprocal-space spin texture, as exemplified by conventional collinear antiferromagnets and altermagnets, respectively. A coplanar or noncoplanar real-space order can likewise produce 0D, 1D, 2D or 3D reciprocal-space textures depending on the full SSG symmetry. This independence between real- and reciprocal-space constraints is one of the central features of the SSG classification \cite{SongUnifiedPRX}.

\subsection{Conventional and unconventional magnetism}

We use the dimensionality of the generic-$\mathbf k$ spin texture to define conventional and unconventional compensated magnetism in the present work. For a zero-net-moment magnetic state, we define
\begin{equation}
    d=0
    \quad\Longrightarrow\quad
    \text{conventional},
    \label{eq:conventional_definition}
\end{equation}
whereas
\begin{equation}
    d>0
    \quad\Longrightarrow\quad
    \text{unconventional}.
    \label{eq:unconventional_definition}
\end{equation}

A conventional compensated magnet therefore has no symmetry-allowed spin polarization at generic momentum. In the simplest collinear case, this corresponds to a conventional antiferromagnet in which the electronic bands remain spin degenerate throughout the Brillouin zone in the non-relativistic limit. The vanishing spin texture can be enforced, for example, by a combined inversion--time-reversal-like symmetry or by other SSG operations whose local constraints eliminate all nonzero components of $\mathbf S(\mathbf k)$.

By contrast, an unconventional compensated magnet possesses a finite-dimensional symmetry-allowed spin texture at generic momentum despite having zero net magnetization in real space. The unconventional class therefore contains not only the familiar spin-polarized altermagnetic and odd-wave states, but also magnetic phases with two- or three-dimensional reciprocal-space spin textures. The definition used here is consequently broader than definitions that identify unconventional magnetism only with momentum-dependent collinear spin splitting.

The classification can be summarized as
\begin{equation}
\begin{array}{ccl}
d=0 &:& \text{conventional},\\[2pt]
d=1 &:& \text{unconventional, spin polarized},\\[2pt]
d=2 &:& \text{unconventional, spin coplanar},\\[2pt]
d=3 &:& \text{unconventional, spin noncoplanar}.
\end{array}
\label{eq:texture_classification_summary}
\end{equation}

This broad definition is used consistently throughout the present work. In particular, unless explicitly stated otherwise, the term ``unconventional'' in the benchmark and Materials Project statistics refers to all candidates with $d>0$. The more restrictive term ``spin polarized'' refers specifically to the $d=1$ subset.

This distinction is important for interpreting the high-throughput results. A material can be predicted to be unconventional even if some or all of its compatible states possess 2D or 3D spin textures and therefore do not belong to the spin-polarized subset. Conversely, a material for which every compatible candidate is spin polarized satisfies a substantially stronger constraint than one for which every candidate is unconventional in the broader sense. This distinction forms the basis of the S1--S6 material-level hierarchy introduced below.

\subsection{Spin-polarized magnetic states}

The $d=1$ class deserves separate consideration because its reciprocal-space spin polarization has the particularly simple form
\begin{equation}
    \mathbf S(\mathbf k)
    =
    s(\mathbf k)\hat{\mathbf n},
    \label{eq:spin_polarized_general}
\end{equation}
where $\hat{\mathbf n}$ is a fixed spin-space direction and $s(\mathbf k)$ is a scalar function of momentum. All symmetry-allowed spin polarization is therefore collinear in reciprocal space, even when the underlying real-space magnetic configuration is coplanar or noncoplanar.

In the present classification, Eq.~\eqref{eq:spin_polarized_general} is the defining condition for a symmetry-enforced spin-polarized state. Importantly, this definition does not require $\hat{\mathbf n}\cdot\hat{\mathbf s}$ to be an exact conserved quantity of the electronic Hamiltonian, nor does it require the expectation value of the spin on an individual Bloch state to be quantized.

It is therefore useful to distinguish the symmetry-defined spin-polarized condition from a stronger spin-conserving limit. If the Hamiltonian additionally satisfies
\begin{equation}
    [\hat H,\hat S_{\hat{\mathbf n}}]=0,
    \label{eq:spin_conservation}
\end{equation}
where
\begin{equation}
    \hat S_{\hat{\mathbf n}}
    =
    \hat{\mathbf n}\cdot\hat{\mathbf S},
\end{equation}
then spin along $\hat{\mathbf n}$ is a good quantum number. Electronic bands can then be separated into two independent spin sectors, and their spin expectation values are quantized along the common spin axis in the ideal non-relativistic limit. Collinear ferromagnets and collinear altermagnets provide familiar examples of this stronger situation.

A general symmetry-enforced 1D texture need not satisfy Eq.~\eqref{eq:spin_conservation}. In this case the SSG constrains only the direction of the allowed expectation value,
\begin{equation}
    \langle\hat{\mathbf S}\rangle_{n\mathbf k}
    =
    s_n(\mathbf k)\hat{\mathbf n},
    \label{eq:nonquantized_spin_polarization}
\end{equation}
while the magnitude $|s_n(\mathbf k)|$ need not be quantized and may vary continuously with momentum and band index. It may even vanish at particular momenta without requiring the bands to become degenerate. Coplanar even-wave magnetic states provide an important example of this more general form of spin polarization \cite{SongUnifiedPRX}.

Thus, the term ``spin polarized'' is used in this work in the symmetry sense of a one-dimensional allowed reciprocal-space spin subspace. The stronger spin-conserving condition is a special subset,
\begin{equation}
    \text{spin-conserving}
    \subset
    \text{1D spin-polarized}.
    \label{eq:spin_conserving_subset}
\end{equation}
This distinction allows collinear altermagnets and more general coplanar spin-polarized phases to be described within the same symmetry framework while retaining their different microscopic spin properties.

\subsection{Wave symmetry of spin-polarized states}

Once the reciprocal-space spin texture is one-dimensional, its remaining momentum dependence can be classified by the scalar polarization function $s(\mathbf k)$ defined through Eq.~\eqref{eq:spin_polarized_general}. An SSG operation maps the common polarization axis either onto itself or onto its opposite. For an operation $g=\{U\Vert R|\boldsymbol{\tau}\}$, we write
\begin{equation}
    U\hat{\mathbf n}=\eta_g\hat{\mathbf n},
    \qquad \eta_g=\pm1.
    \label{eq:wave_axis_character}
\end{equation}
The scalar polarization then satisfies
\begin{equation}
    s(\mathbf k)
    =\eta_g\,s\!\left[\det(U)R\mathbf k\right],
    \label{eq:scalar_spin_transformation}
\end{equation}
The factor $\det(U)$ explicitly includes the momentum reversal associated with antiunitary operations, consistently with Eq.~\eqref{eq:ssg_reciprocal_space_invariance}. Spin-preserving and spin-flipping operations carry characters $\eta_g=+1$ and $-1$, respectively.

The wave symmetry is obtained by finding momentum-space basis functions satisfying the same transformation law. We denote homogeneous polynomial basis functions of total degree $n$ by $F_\alpha^{(n)}(\mathbf k)$, where $\alpha$ distinguishes linearly independent basis functions at the same degree. The basis index is omitted when only one function is needed. In a monomial representation,
\begin{equation}
    F_\alpha^{(n)}(\mathbf k)
    =\sum_{i+j+m=n}c_{\alpha,ijm}k_x^i k_y^j k_z^m,
    \label{eq:wave_polynomial}
\end{equation}
where $n,i,j,m\in\mathbb Z_{\ge0}$. Parenthesized superscripts denote polynomial degree throughout; wave labels are stated separately. At each degree, the coefficients are constrained by
\begin{equation}
    F_\alpha^{(n)}(\mathbf k)
    =\eta_g F_\alpha^{(n)}\!\left[\det(U)R\mathbf k\right]
    \qquad\text{for all }g\in\mathcal G^S.
    \label{eq:wave_polynomial_constraint}
\end{equation}
A nonzero solution defines an allowed basis function. The lowest allowed degree in a given parity sector determines its leading wave label. The wave label specifies a symmetry-allowed polynomial order and does not require the full spin texture throughout the Brillouin zone to equal a single homogeneous polynomial. The wave label characterizes the lowest-degree polynomial basis functions allowed by symmetry. It does not, by itself, assert that the spin expectation value of an individual Bloch state admits a Taylor expansion with that leading degree.

The distinction between odd-, even- and mixed-wave textures follows by resolving the scalar polarization into its two parity components,
\begin{equation}
    s(\mathbf k)=s_{\mathrm{odd}}(\mathbf k)+s_{\mathrm{even}}(\mathbf k),
    \qquad
    s_{\mathrm{odd/even}}(\mathbf k)
    =\frac{s(\mathbf k)\mp s(-\mathbf k)}{2}.
    \label{eq:wave_parity_decomposition}
\end{equation}
These components satisfy
\begin{equation}
    s_{\mathrm{odd}}(-\mathbf k)=-s_{\mathrm{odd}}(\mathbf k),
    \qquad
    s_{\mathrm{even}}(-\mathbf k)=s_{\mathrm{even}}(\mathbf k).
    \label{eq:wave_parity_relations}
\end{equation}
Because momentum reversal commutes with the linear momentum transformations in Eq.~\eqref{eq:scalar_spin_transformation}, the polynomial constraints can be solved separately in the odd and even sectors. We denote their lowest allowed degrees by $\ell_{\mathrm{odd}}$ and $\ell_{\mathrm{even}}$, respectively.

For an odd-wave texture, symmetry permits only the odd component, and the leading labels include $p$-, $f$-, $h$-, $k$- and $m$-wave for degrees 1, 3, 5, 7 and 9. For an even-wave texture, symmetry permits only the even component; the leading even-wave labels in the present compensated SSG classification are $d$-, $g$- and $i$-wave for degrees 2, 4 and 6. Representative basis functions are $k_x$ for $p$ wave, $k_xk_y$ or $k_x^2-k_y^2$ for $d$ wave, and $k_xk_yk_z$ for $f$ wave. Their precise form is determined by the SSG constraints. These labels refer to the leading allowed orders; higher-order contributions of the same parity can also be present.

When both parity sectors are allowed, the one-dimensional texture has mixed-wave symmetry. We use an additive label formed from the lowest allowed order in each sector, with the odd-wave label preceding the even-wave label. A $p+d$ label, for example, denotes odd and even contributions beginning at degrees 1 and 2 within the same scalar polarization,
\begin{equation}
    \mathbf S(\mathbf k)
    =\left[a F^{(1)}(\mathbf k)+b F^{(2)}(\mathbf k)+\cdots\right]\hat{\mathbf n}.
    \label{eq:mixed_wave_notation}
\end{equation}
Here $F^{(1)}$ and $F^{(2)}$ are real basis functions compatible with the SSG, $a$ and $b$ are real amplitudes, and the ellipsis denotes higher-order contributions. The label specifies the allowed wave orders without fixing their amplitudes or relative signs. Likewise, $f+g$ denotes $\ell_{\mathrm{odd}}=3$ and $\ell_{\mathrm{even}}=4$, whereas $f+i$ denotes $\ell_{\mathrm{odd}}=3$ and $\ell_{\mathrm{even}}=6$. Both contributions belong to the same scalar polarization along a fixed spin axis, so their coexistence does not increase the spin-texture dimension.

For collinear compensated magnets, the spin-only symmetry enforces an even reciprocal-space polarization, so the spin-polarized unconventional states correspond to even-wave altermagnets. For coplanar magnetic orders, the parity depends on the orientation of the reciprocal-space polarization relative to the real-space spin plane: a perpendicular polarization is odd under $\mathbf k\rightarrow-\mathbf k$, whereas an in-plane polarization is even \cite{SongUnifiedPRX}.

Noncoplanar magnetic orders lack the nontrivial spin-only symmetries that fix parity in collinear and coplanar states. Their parity is instead determined by the combined spin-space operations of the full SSG. A one-dimensional texture can therefore be odd-wave, even-wave or mixed-wave, depending on whether symmetry allows the odd sector, the even sector or both. In the mixed-wave case, the odd and even contributions share the same fixed polarization axis. Noncoplanarity thus permits mixed parity without requiring it. The symmetry origin of mixed-wave magnetism and its connection to the magnetic candidates of \ch{VGe3} are discussed in Section~\ref{sec:mixed_wave_magnetism}.

\subsection{Wave symmetry of multidimensional spin textures}

Two- and three-dimensional textures can combine different wave symmetries across spin components, termed hybrid-wave\cite{Luo2026Unconventional,Ryu2026MixedParityTetraborides}. In a specified spin coordinate system, choosing orthonormal axes $\hat{\mathbf e}_a$ spanning the symmetry-allowed spin subspace, we write
\begin{equation}
    \mathbf S(\mathbf k)=\sum_{a=1}^{d}s_a(\mathbf k)\hat{\mathbf e}_a,
    \qquad d=2,3.
    \label{eq:multidimensional_wave_components}
\end{equation}
Each component can be resolved into momentum-space basis functions and characterized by its leading allowed wave orders. The constraints are obtained from the full vector transformation law,
\begin{equation}
    s_a\!\left[\det(U)R\mathbf k\right]
    =\sum_{b=1}^{d}D_{ab}(g)\,s_b(\mathbf k),
    \qquad
    D_{ab}(g)=\hat{\mathbf e}_a\cdot U\hat{\mathbf e}_b.
    \label{eq:multidimensional_wave_constraint}
\end{equation}
SSG operations can mix spin components, so their polynomial coefficients must satisfy these vector constraints jointly. The wave content can then be read from the projections onto the specified axes; an individual component need not transform as an independent one-dimensional representation.

At each degree, we denote a vector basis function satisfying these constraints by $\mathbf F_\alpha^{(n)}(\mathbf k)$ and its spin components by $F_{a,\alpha}^{(n)}(\mathbf k)$,
\begin{equation}
    \mathbf F_\alpha^{(n)}(\mathbf k)
    =\sum_{a=1}^{d}F_{a,\alpha}^{(n)}(\mathbf k)\hat{\mathbf e}_a.
    \label{eq:vector_wave_basis_notation}
\end{equation}
Here $a$ labels the spin component and $\alpha$ labels the vector basis function. When the basis index is omitted, component functions are written as $F_x^{(n)}$, $F_y^{(n)}$ or $F_z^{(n)}$.

Hybrid-wave labels form an ordered tuple in the specified spin coordinates. For example, $(d,f)_{xy}$ denotes a two-dimensional texture with leading $d$- and $f$-wave components along $x$ and $y$,
\begin{equation}
    \mathbf S(\mathbf k)
    =A_d F_x^{(2)}(\mathbf k)\hat{\mathbf x}_s
    +A_f F_y^{(3)}(\mathbf k)\hat{\mathbf y}_s
    +\cdots.
    \label{eq:hybrid_wave_example}
\end{equation}
Here the basis functions and their amplitudes satisfy the full vector symmetry constraints. The omitted $z$ component vanishes identically in this coordinate system. For a three-dimensional texture, the notation $(p,d,d)_{xyz}$ denotes
\begin{equation}
    \mathbf S(\mathbf k)
    =a F_x^{(1)}(\mathbf k)\hat{\mathbf x}_s
    +b F_y^{(2)}(\mathbf k)\hat{\mathbf y}_s
    +c F_z^{(2)}(\mathbf k)\hat{\mathbf z}_s
    +\cdots,
    \label{eq:hybrid_wave_3d_notation}
\end{equation}
where the subscripts identify the spin components and the parenthesized superscripts give the polynomial degrees. The two $d$ entries specify the same leading wave order without requiring identical basis functions or amplitudes. These tuples summarize the component wave content; the SSG can relate the coefficients through Eq.~\eqref{eq:multidimensional_wave_constraint}. Rotating the spin coordinate system can mix the components, so the axes must be specified. A two-dimensional texture can have three nonzero Cartesian projections if its spin plane is tilted relative to the chosen axes; its tuple then retains all three entries, while its spin-texture dimension remains two.

Each component is resolved into odd and even sectors using the same wave-order convention as for a one-dimensional texture. An entry can therefore itself carry an additive mixed-wave label: $(p+d,f,d)_{xyz}$ denotes a mixed $p+d$ component along $x$, an $f$-wave component along $y$ and a $d$-wave component along $z$. Mixed-wave thus describes odd and even contributions within one scalar component, whereas hybrid-wave describes wave symmetries across different spin components. All spin components are real expectation values. A zero of the full vector requires every component to vanish, while contributions within one component can cancel. Neither notation changes the dimension of the symmetry-allowed spin subspace.

The candidate-level characterization can therefore be organized by real-space geometry, reciprocal-space dimension and wave symmetry. For $d=1$, the wave symmetry is specified by the allowed odd and even sectors and their leading orders. For $d=2,3$, it can be examined through the component basis functions subject to the full vector constraints. The automated high-throughput tables used here assign wave labels to the $d=1$ candidates, including mixed-wave states; $d=2,3$ candidates retain their texture-dimensionality labels. A complete enumeration of their component wave symmetries is not part of the present tables.

\begin{equation}
    \text{real-space geometry}
    \;+\;d
    \;+\;\text{component wave symmetry}.
    \label{eq:candidate_classification}
\end{equation}
The conventional, unconventional and spin-polarized categories remain defined by $d=0$, $d>0$ and $d=1$, respectively. In particular, mixed-wave candidates remain spin polarized and do not change the S1--S6 assignment rules. Section~\ref{sec:prediction_workflow} describes how the candidate classifications enter the high-throughput prediction workflow.

%################################################################################
%##### SECTION: Complete symmetry-based prediction workflow
%################################################################################

\section{Spin-translation symmetry and the conserved hyperspin component}
\label{sec:hyperspin_spin_translation}

Fixed-axis spin polarization follows directly from spin-translation symmetry. For an operation $g_\theta=\{C_{\hat{\mathbf n}}(\theta)\Vert E|\boldsymbol\tau\}$, its representation in orbital and physical-spin space is
\begin{equation}
 D_{\mathbf k}(g_\theta)
 =T_{\boldsymbol\tau}(\mathbf k)\otimes e^{-i\theta\hat S_{\hat{\mathbf n}}/\hbar},
 \qquad [H(\mathbf k),D_{\mathbf k}(g_\theta)]=0.
 \label{eq:hyperspin_spin_translation}
\end{equation}
A translation leaves momentum unchanged and commutes with physical spin. Choosing $\hat{\mathbf n}=\hat{\mathbf z}$ therefore gives
\begin{equation}
 D_{\mathbf k}(g_\theta)\hat S_+D_{\mathbf k}(g_\theta)^{-1}
 =e^{-i\theta}\hat S_+,
 \qquad \hat S_+=\hat S_x+i\hat S_y.
 \label{eq:hyperspin_transverse_constraint}
\end{equation}
For a nondegenerate energy eigenstate, this requires $\langle\hat S_x\rangle=\langle\hat S_y\rangle=0$ whenever $\theta\not\equiv0\pmod{2\pi}$. The same result holds for simultaneous symmetry eigenstates within a degenerate subspace. This constraint applies throughout the Brillouin zone, independently of the sublattice basis, and does not require $[H,\hat S_z]=0$. General $C_n$ operations provide conserved symmetry labels, which need not be two-valued.

We first consider the two-sublattice $p$-wave model\cite{hellenes2024pwavemagnets},
\begin{equation}
 H_{\mathrm p}=2t\left[\cos(k_x/2)\tau_x+\cos k_y\mathbb{I}\right]
 +2t_J\left[\sin(k_x/2)\sigma_x\tau_y+\cos k_y\sigma_y\tau_z\right].
 \label{eq:hyperspin_pwave_model}
\end{equation}
Here $\tau_i$ and $\sigma_i$ act on sublattice and physical spin, respectively, whereas $\boldsymbol\tau$ denotes a translation vector. The half translation exchanges the two sites. In a Bloch convention including their positions, the operation $g_\pi=\{C_{2z}\Vert E|\boldsymbol\tau\}$ is represented by
\begin{equation}
 \begin{aligned}
 T_{\boldsymbol\tau}(\mathbf k)&=e^{-ik_x/2}\tau_x,\qquad
 D_{\mathbf k}(g_\pi)=-i e^{-ik_x/2}\tau_x\sigma_z,\\
 Q_g&\equiv i e^{ik_x/2}D_{\mathbf k}(g_\pi)=\tau_x\sigma_z,
 \qquad J_\parallel=\frac{\hbar}{2}Q_g.
 \end{aligned}
 \label{eq:hyperspin_normalized_operation}
\end{equation}
Each term in $H_{\mathrm p}$ commutes with $Q_g$, which is Hermitian and satisfies $Q_g^2=\mathbb{I}$. Its eigenvalues $\pm1$ label combined sublattice--spin sectors; they do not fix the magnitude of the physical spin expectation value.

The hyperspin formulation of Ref.~\cite{Ma2026HyperspinAltermagnets} encompasses this $p$-wave physics in a different, subsystem-based Hamiltonian; its broader construction also admits higher-order odd-wave states. The model retains a pair of states $|\mu,+,\mathbf k\rangle$, $|\mu,-,\mathbf k\rangle$ from each AFM subsystem $\mu=A,B$. Its assumed $\Theta T_{\boldsymbol\tau}$ symmetry, combined with the coplanar spin-only symmetry $C_{2y}\Theta$, gives $g=\{C_{2y}\Vert E|\boldsymbol\tau\}$ up to an overall spinor phase. Here $\Theta$ denotes time reversal and $y$ is normal to the spin plane. Using symmetry-adapted relative phases for the subsystem pairs, the action is
\begin{equation}
 \begin{aligned}
 g|\mu,+,\mathbf k\rangle
 &=e^{-i\mathbf k\cdot\boldsymbol\tau}|\mu,-,\mathbf k\rangle,\\
 g|\mu,-,\mathbf k\rangle
 &=-e^{-i\mathbf k\cdot\boldsymbol\tau}|\mu,+,\mathbf k\rangle.
 \end{aligned}
 \label{eq:hyperspin_subsystem_action}
\end{equation}
The first relation fixes the paired-state phase; the second follows from $D_{\mathbf k}(g)^2=-e^{-2i\mathbf k\cdot\boldsymbol\tau}\mathbb{I}$. Translation acts within each subsystem, and its internal-site action is included in these full Bloch states. Thus, in the ordered basis $(A+,A-,B+,B-)$,
\begin{equation}
 \begin{aligned}
 D_{\mathbf k}(g)&=-i e^{-i\mathbf k\cdot\boldsymbol\tau}
 \begin{pmatrix}\nu_y&0\\0&\nu_y\end{pmatrix},\\
 Q_g&\equiv i e^{i\mathbf k\cdot\boldsymbol\tau}D_{\mathbf k}(g)
 =\begin{pmatrix}\nu_y&0\\0&\nu_y\end{pmatrix}
 =\frac{2J_y}{\hbar}.
 \end{aligned}
 \label{eq:hyperspin_direct_identification}
\end{equation}
Here $\nu_y$ acts on each retained pair, and the last equality uses the hyperspin matrix of Ref.~\cite{Ma2026HyperspinAltermagnets} in this paired-state convention. It is not the bare physical-spin matrix in an explicit site basis. The Hamiltonian in that representation reads
\begin{equation}
 H_{\mathrm h}=
 \begin{pmatrix}
 (h_0+h_1)\mathbb{I}+h_4\nu_y&h_2\mathbb{I}+h_3\nu_y\\
 h_2^*\mathbb{I}+h_3^*\nu_y&(h_0-h_1)\mathbb{I}+h_5\nu_y
 \end{pmatrix},
 \label{eq:hyperspin_hamiltonian_blocks}
\end{equation}
whose blocks all commute with $\nu_y$. Consequently, $[H_{\mathrm h},Q_g]=[H_{\mathrm h},J_y]=0$: the conserved hyperspin component is represented by the normalized spin-translation operation in this symmetry-adapted basis.

Finally, the coplanar $d$-wave model of Ref.~\cite{SongUnifiedPRX} realizes the same conserved structure without $\Theta T_{\boldsymbol\tau}$. Its Hamiltonian is
\begin{equation}
 \begin{aligned}
 H_{\mathrm d}&=\varepsilon\mathbb{I}+A\tau_x+D\sigma_x\tau_z
                  +B\sigma_x\tau_y+C\sigma_y\tau_z,\\
 \varepsilon&=2t_0\cos k_y,\qquad A=2t_0\cos(k_x/2),\\
 B&=t_3\sin(k_x/2),\qquad C=t_4\sin k_y,\\
 D&=t_1\cos k_x+t_2\cos k_y.
 \end{aligned}
 \label{eq:hyperspin_dwave_model}
\end{equation}
The half translation again exchanges the two effective sites, giving $D_{\mathbf k}(g)=-i e^{-ik_x/2}\tau_x\sigma_z$ for $g=\{C_{2z}\Vert E|\boldsymbol\tau\}$. Every Hamiltonian term commutes with $Q_g=\tau_x\sigma_z$, so $J_\parallel^{(\mathrm d)}=\hbar Q_g/2$ defines a conserved hyperspin-like component even though $[H_{\mathrm d},\sigma_z]$ is generally nonzero. For the same half translation, however, the matrix part of $\Theta T_{\boldsymbol\tau}$ is proportional to $\tau_x\sigma_y$, and
\begin{equation}
 (\tau_x\sigma_y)H_{\mathrm d}(-\mathbf k)^*(\tau_x\sigma_y)
 =H_{\mathrm d}(\mathbf k)-2C\sigma_y\tau_z.
 \label{eq:hyperspin_absent_time_translation}
\end{equation}
Thus $\Theta T_{\boldsymbol\tau}$ is broken for generic $t_4\ne0$, whereas $C_{2z}T_{\boldsymbol\tau}$ remains a symmetry. This even-wave model therefore supports a conserved composite spin label without time reversal combined with translation. The protecting operation is the unitary spin translation; additional SSG operations determine the alternating momentum dependence and its parity. Consequently, $\Theta T_{\boldsymbol\tau}$ is not a necessary condition for this conservation mechanism or for the fixed-axis alternating spin splitting realized in the model.

\section{Complete symmetry-based prediction workflow}
\label{sec:prediction_workflow}

The complete prediction workflow starts from a non-magnetic crystallographic structure together with a specified magnetic sublattice and constructs all zero-net-moment magnetic candidates contained within a controlled SSG search space. Each retained candidate is then classified according to its real-space magnetic geometry and reciprocal-space spin texture, and the resulting candidate-level information is finally aggregated at the material level. The workflow can therefore be summarized as
\begin{equation}
\begin{split}
    &\text{crystallographic structure}
    +\text{ magnetic-sublattice information}
    \\
    &\qquad\Downarrow
    \\
    &\text{identification of the magnetic sublattice}
    \\
    &\qquad\Downarrow
    \\
    &\text{crystallographic standardization}
    \\
    &\qquad\Downarrow
    \\
    &\text{enumeration and construction of compatible SSG candidates}
    \\
    &\qquad\Downarrow
    \\
    &\text{candidate-level magnetic classification}
    \\
    &\qquad\Downarrow
    \\
    &\text{material-level candidate ensemble}.
\end{split}
\label{eq:workflow_overview}
\end{equation}

Within the present implementation, the search is restricted to compensated magnetic states with vanishing net magnetization, one magnetic species occupying a single symmetry-equivalent crystallographic orbit, and commensurate SSG candidates with $I_k\leq4$. These conditions define the domain of applicability of the workflow. The detailed procedures used to obtain the benchmark and Materials Project input datasets are described separately in the corresponding sections below.

\subsection{Input and applicability conditions}

The magnetic classification developed in this work is designed for compensated magnetic states with vanishing net magnetization. Accordingly, the basic physical applicability condition is that the material can support a zero-net-moment magnetic state. For an experimentally studied material, whether the net magnetization is zero or negligibly small can generally be established much more readily than the complete magnetic configuration, for example from bulk magnetization or related magnetic measurements. Materials with a robust finite ferromagnetic or ferrimagnetic moment therefore lie outside the direct target of the present classification.

Apart from the crystallographic structure itself, the only magnetic information required as input is the identification of the magnetic sublattice, namely which atomic species and sites are expected to carry magnetic moments. The detailed moment directions, propagation vector and magnetic ground-state SSG are not required at this stage. How this minimal input information is extracted from MAGNDATA and the Materials Project is described in the benchmark and high-throughput sections, respectively.

\subsection{Step 1---Identification of the magnetic sublattice}

The first step identifies whether the magnetic sublattice satisfies the structural conditions required by the present SSG construction. We restrict the workflow to one magnetic chemical species occupying a single symmetry-equivalent crystallographic orbit.

The single-orbit condition is physically natural for the present construction. Atoms of the same chemical element located at different Wyckoff positions generally experience different local chemical environments and can therefore possess different magnetic moments or magnetic responses. Such sites are effectively independent magnetic sublattices and are not treated as a single equivalent magnetic species in the present workflow.

Let the magnetic sites be
\begin{equation}
    \{\mathbf r_1,\mathbf r_2,\ldots,\mathbf r_N\}.
\end{equation}
They are required to form a single orbit under the crystallographic space group, such that for any two magnetic sites $\mathbf r_i$ and $\mathbf r_j$ there exists a spatial symmetry operation satisfying
\begin{equation}
    \mathbf r_j
    =
    \{R|\boldsymbol{\tau}\}\mathbf r_i
\end{equation}
up to a lattice translation.

The retained input structures therefore satisfy the following conditions:
\begin{enumerate}
    \item the material is considered within the zero-net-moment magnetic sector;
    \item exactly one chemical species forms the magnetic sublattice;
    \item all magnetic sites of that species belong to one symmetry-equivalent crystallographic orbit.
\end{enumerate}

These conditions ensure that the magnetic configuration can be generated from a single reference magnetic site using the SSG operations. Because all magnetic sites belong to the same symmetry orbit, the symmetry construction also relates their moment amplitudes and orientations. Materials containing multiple independent magnetic species or multiple inequivalent magnetic Wyckoff orbits are outside the present implementation and are discussed separately in the section on scope and limitations.

\subsection{Step 2---Crystallographic standardization}

The SSG construction requires the crystallographic structure to be expressed in a setting that is consistent with the spatial operations used in the SSG database. Each retained input structure is therefore standardized before the compatible SSGs are enumerated.

The non-magnetic crystallographic structure is analysed using \texttt{spglib} to determine its space group, standardized lattice, atomic coordinates and symmetry-equivalent Wyckoff positions. The structure is then transformed to the standard crystallographic setting used throughout the subsequent SSG construction. After this standardization, the full set of spatial symmetry operations of the crystal can be directly compared with the spatial projection of the enumerated SSGs.

This step is necessary because crystallographically equivalent structures may be represented using different primitive cells, origin choices or coordinate settings. Although such descriptions are physically equivalent, a unique standardized representation is required for an unambiguous correspondence between the input crystal structure and the spatial operations contained in the SSG database.

The output of this step is therefore a standardized non-magnetic crystal structure with a well-defined space group and a uniquely identified magnetic Wyckoff position. This standardized structure provides the input for the enumeration and explicit construction of compatible SSG magnetic candidates.

\subsection{Step 3---Enumeration of compatible zero-net-moment SSGs}

This step constitutes the central symmetry-construction stage of the workflow. For each standardized crystal structure, we first enumerate SSGs compatible with the crystallographic symmetry and the chosen magnetic-cell range, and then construct the corresponding magnetic configurations on the specified magnetic Wyckoff position.

\paragraph{Parent-SG matching.}

The first requirement is that the parent SG of the candidate SSG coincides with the space group of the non-magnetic crystallographic structure, $G_{\mathrm{parent}}^{\mathrm{SSG}}=G_{\mathrm{crystal}}$. As defined above, the parent SG is obtained by projecting the full SSG onto real space and removing the spin-space part of each operation. The present construction therefore retains the full spatial symmetry of the non-magnetic crystal. Magnetic orders whose SSG parent SG is a proper subgroup of the crystallographic space group are outside the search space considered in this work.

\paragraph{Magnetic-cell cutoff.}

Among all SSGs with the required parent SG, we retain only those with $I_k\leq4$. As discussed above, this range covers nearly all experimentally reported magnetic structures in the MAGNDATA dataset considered here and includes the dominant experimentally observed cases with $I_k=1$ and $I_k=2$.

\paragraph{Zero-net-moment symmetry screening.}

The unconventional-magnetism classification considered in this work is defined for compensated magnetic states. We therefore retain only SSGs whose spin-space symmetry forbids a nonzero net magnetization. Using the criterion introduced in Section~\ref{sec:ssg_classification}, this corresponds to a nonpolar spin-part point group, for which no nonzero vector is invariant under all spin-space operations. Consequently, any magnetic structure realizing such an SSG has $\sum_i\mathbf m_i=0$.

This symmetry condition directly implements the physical domain of the present workflow. For an experimentally studied material, whether a substantial net magnetization is present can usually be established by comparatively simple magnetic measurements before the detailed magnetic structure is known. Ferromagnetic or ferrimagnetic materials with a finite net moment therefore do not enter the unconventional compensated-magnet classification considered here.

\paragraph{Reference-site construction.}

For each remaining SSG, one site $\mathbf r_0$ is selected from the magnetic Wyckoff position and assigned a reference magnetic moment $\mathbf m_0$. Its allowed direction is determined by the onsite SSG operations. We define
\begin{equation}
    G_{\mathbf r_0}
    =
    \left\{
    \{U_\alpha\Vert R_\alpha|\boldsymbol{\tau}_\alpha\}
    \;\middle|\;
    \{R_\alpha|\boldsymbol{\tau}_\alpha\}\mathbf r_0
    =
    \mathbf r_0+\mathbf R_n
    \right\},
    \label{eq:onsite_ssg_group}
\end{equation}
where $\mathbf R_n$ is a lattice translation. The reference moment must satisfy
\begin{equation}
    U_\alpha\mathbf m_0=\mathbf m_0,
    \qquad
    \forall\,
    \{U_\alpha\Vert R_\alpha|\boldsymbol{\tau}_\alpha\}
    \in G_{\mathbf r_0}.
    \label{eq:reference_moment_constraint}
\end{equation}

The common invariant subspace of these onsite spin operations determines the allowed magnetic degrees of freedom at the reference site. If this subspace is zero-dimensional, no nonzero magnetic moment is compatible with the target SSG on the specified magnetic Wyckoff position and the candidate is discarded. If the allowed subspace is one-dimensional, the moment direction is fixed up to its magnitude and sign. For a two- or three-dimensional allowed subspace, a representative vector is formed using fixed nonzero coefficients in a basis of that subspace and then normalized.

The particular choice of this generic vector does not affect the symmetry-based classification because all such choices satisfy the same SSG constraints. Its role is only to provide an explicit representative magnetic configuration. Special choices can, however, accidentally introduce additional symmetry; such cases are identified by the full-SSG verification described below.

\paragraph{Generation of the full magnetic configuration.}

Once the reference moment has been chosen, the moments on all other symmetry-equivalent sites belonging to the same magnetic Wyckoff position are generated directly by the SSG operations. If an operation relates the reference site to a site $\mathbf r_i$ according to
\begin{equation}
    \mathbf r_i
    =
    \{R_\alpha|\boldsymbol{\tau}_\alpha\}\mathbf r_0,
\end{equation}
the corresponding magnetic moment is
\begin{equation}
    \mathbf m_i
    =
    U_\alpha\mathbf m_0.
    \label{eq:generated_moment}
\end{equation}

Because all magnetic sites belong to a single Wyckoff position and the reference moment has already been constrained by its complete onsite subgroup, Eq.~\eqref{eq:generated_moment} consistently determines the magnetic moments on all symmetry-equivalent sites. The absolute amplitude of $\mathbf m_0$ is irrelevant for the symmetry construction; only the relative orientations generated by the SSG are required. A normalized reference moment can therefore be used without loss of generality.

\paragraph{Full-SSG verification.}

The magnetic configuration constructed above is guaranteed to possess all operations of the target SSG because it is generated explicitly from those operations. It is not, however, guaranteed that the target SSG is the full symmetry of the resulting configuration. A particular choice of the reference moment may accidentally introduce additional symmetry. For example, a generic coplanar SSG can yield a collinear configuration if the selected moment happens to lie along a special direction, thereby introducing additional spin-only operations.

We therefore identify the full SSG of each generated magnetic configuration and retain the candidate only when its realized SSG coincides with the target SSG.
The purpose of this step is not to verify that the target operations are present, which follows directly from the construction, but to ensure that no additional symmetry has been introduced accidentally.

In the present work, the parent SG of every candidate is required to coincide with the full space group of the non-magnetic crystal. The atomic structure therefore cannot acquire an additional spatial operation beyond those already contained in the parent SG. Any accidental enlargement of the generated SSG can consequently arise only through additional spin-only symmetry. In a more general construction in which the candidate SSG has a parent SG that is a proper subgroup of the non-magnetic crystallographic space group, additional operations involving nontrivial spatial transformations would also have to be examined; such cases are not included in the present workflow.

\subsection{Step 4---Classification of retained SSG candidates}

Each retained magnetic structure is classified according to its realized SSG using the symmetry criteria introduced in Section~\ref{sec:ssg_classification}. Because the classification is determined entirely by the SSG, magnetic structures realizing the same SSG are assigned the same magnetic class.

For each candidate, we record both its real-space magnetic geometry and its reciprocal-space spin-texture classification. The real-space configuration is classified as collinear, coplanar or noncoplanar according to its spin-only symmetry. In reciprocal space, the symmetry-allowed spin texture is classified by its dimension $d=0,1,2,3$. Candidates with $d=0$ are classified as conventional, whereas those with $d>0$ are classified as unconventional. The $d=1$ subset corresponds to spin-polarized states and is further characterized by its odd-, even- or mixed-wave symmetry. For noncoplanar spin-1D candidates, the lowest allowed odd and even orders are recorded separately. Candidates with $d=2,3$ retain their dimensionality labels in the current tables; their component wave symmetries, including possible hybrid-wave textures, can be analysed using the vector constraints described above.

In the present prediction framework, the reciprocal-space classification is of primary interest because it directly distinguishes conventional from unconventional magnetic character. We therefore associate each material with the distribution of conventional, unconventional and spin-polarized states among all of its compatible SSG candidates. These candidate-level classifications form the direct input to the material-level prediction hierarchy introduced in the following section.

\subsection{Step 5---Material-level aggregation}

The final step combines the classifications of all compatible SSG candidates associated with the same material. Rather than selecting a single magnetic configuration, we retain the complete candidate ensemble and record how many candidates are conventional, unconventional and spin polarized, together with their magnetic-cell indices and, where applicable, wave symmetries.

This material-level aggregation provides the information required to evaluate how strongly the crystallographic structure constrains the magnetic character of a material. The resulting candidate ensembles are then organized into the S1--S6 prediction hierarchy introduced in the following section.

%################################################################################
%##### SECTION: Material-level prediction hierarchy
%################################################################################

\section{Material-level prediction hierarchy}

\subsection{Candidate sets}

The symmetry-based workflow described above produces, for each material, an ensemble of compatible zero-net-moment magnetic candidates rather than a single magnetic configuration. The material-level prediction is therefore formulated in terms of the composition of this candidate ensemble.

For a material $\mathcal{M}$, we denote the complete set of compatible candidates within the present magnetic-cell cutoff as
\begin{equation}
    C_{\leq4}(\mathcal{M})
    =
    \left\{
    c\,\middle|\,I_k(c)\leq4
    \right\},
    \label{eq:candidate_set_le4}
\end{equation}
where each candidate $c$ has passed the symmetry construction and full-SSG verification described above. We further define the subset that does not enlarge the crystallographic primitive cell as
\begin{equation}
    C_1(\mathcal{M})
    =
    \left\{
    c\in C_{\leq4}(\mathcal{M})
    \,\middle|\,
    I_k(c)=1
    \right\}.
    \label{eq:candidate_set_ik1}
\end{equation}

Each candidate carries the reciprocal-space classification introduced above. A candidate with $d(c)=0$ is conventional, one with $d(c)>0$ is unconventional, and the $d(c)=1$ subset corresponds to spin-polarized states. For convenience, we denote the unconventional condition by $\mathcal U(c)$ and the spin-polarized condition by $\mathcal S(c)$,
\begin{equation}
    \mathcal U(c): d(c)>0,
    \qquad
    \mathcal S(c): d(c)=1.
    \label{eq:US_predicates}
\end{equation}

The material-level prediction is based on whether these properties are shared by all candidates in $C_{\leq4}$ or, when additional magnetic-cell information is considered, by all candidates in $C_1$. The S1--S6 hierarchy is defined from this ensemble-level information.

\subsection{Formal S1--S6 definitions}

We define six mutually exclusive material-level classes according to the composition of $C_{\leq4}$ and, where relevant, its $I_k=1$ subset $C_1$.

A material belongs to S1 when every compatible candidate in the full $I_k\leq4$ search is spin polarized,
\begin{equation}
    \mathrm{S1}:
    \qquad
    \forall c\in C_{\leq4},
    \quad
    \mathcal S(c).
    \label{eq:S1_definition}
\end{equation}
Thus, every compatible candidate satisfies $d(c)=1$.

A material belongs to S2 when every candidate in $C_{\leq4}$ is unconventional, while the S1 condition is not satisfied,
\begin{equation}
    \mathrm{S2}:
    \qquad
    \forall c\in C_{\leq4},
    \quad
    \mathcal U(c),
    \qquad
    \mathrm{not\ S1}.
    \label{eq:S2_definition}
\end{equation}
S2 therefore contains only unconventional candidates but allows 1D, 2D and 3D reciprocal-space spin textures.

For materials that satisfy neither S1 nor S2, we next consider the $I_k=1$ subset. A material belongs to S3 when every candidate in $C_1$ is spin polarized,
\begin{equation}
    \mathrm{S3}:
    \qquad
    \forall c\in C_1,
    \quad
    \mathcal S(c),
    \qquad
    \mathrm{not\ S1,S2}.
    \label{eq:S3_definition}
\end{equation}

Similarly, a material belongs to S4 when every candidate in $C_1$ is unconventional, while the S3 condition is not satisfied,
\begin{equation}
    \mathrm{S4}:
    \qquad
    \forall c\in C_1,
    \quad
    \mathcal U(c),
    \qquad
    \mathrm{not\ S1,S2,S3}.
    \label{eq:S4_definition}
\end{equation}

S5 contains the remaining materials for which conventional and unconventional candidates coexist in $C_{\leq4}$,
\begin{equation}
    \mathrm{S5}:
    \qquad
    \exists\,c_a,c_b\in C_{\leq4}
    \quad
    \mathrm{with}
    \quad
    d(c_a)=0,\;
    d(c_b)>0.
    \label{eq:S5_definition}
\end{equation}

Finally, S6 contains materials for which every compatible candidate in the present search space is conventional,
\begin{equation}
    \mathrm{S6}:
    \qquad
    \forall c\in C_{\leq4},
    \quad
    d(c)=0.
    \label{eq:S6_definition}
\end{equation}

The definitions are summarized in Table~\ref{tab:S1_S6_definition}. The classes are assigned sequentially so that each material receives a unique label.

\begin{table}[htbp]
    \centering
    \caption{Definition and interpretation of the S1--S6 material-level prediction classes. Here $C_{\leq4}$ denotes the complete compatible candidate set with $I_k\leq4$, and $C_1$ denotes its $I_k=1$ subset.}
    \label{tab:S1_S6_definition}
    \begin{tabularx}{\linewidth}{c>{\raggedright\arraybackslash}X>{\raggedright\arraybackslash}X}
        \hline
        Class & Criterion & Interpretation \\
        \hline
        S1 &
        All $C_{\leq4}$ candidates are spin polarized &
        Full candidate set is restricted to $d=1$ \\
        S2 &
        All $C_{\leq4}$ candidates are unconventional &
        Full candidate set is restricted to $d>0$ \\
        S3 &
        All $C_1$ candidates are spin polarized &
        $I_k=1$ subset is restricted to $d=1$ \\
        S4 &
        All $C_1$ candidates are unconventional &
        $I_k=1$ subset is restricted to $d>0$ \\
        S5 &
        Conventional and unconventional candidates coexist &
        Mixed candidate set \\
        S6 &
        All $C_{\leq4}$ candidates are conventional &
        No unconventional candidate within the search space \\
        \hline
    \end{tabularx}
\end{table}

\subsection{Interpretation of the hierarchy}

The S1--S6 classes quantify how strongly the symmetry-compatible candidate ensemble constrains the reciprocal-space magnetic character of a material. They should not be interpreted as statistical probabilities, confidence scores or a universal monotonic ranking.

S1 and S2 are determined from the complete $I_k\leq4$ candidate set and therefore require no additional information about the magnetic-cell size. For an S1 material, every compatible candidate is spin polarized, whereas for an S2 material every compatible candidate is unconventional in the broader sense.

S3 and S4 describe materials for which the full $I_k\leq4$ candidate ensemble does not provide the same level of selectivity, but the ambiguity is removed when the magnetic structure is restricted to $I_k=1$. S3 then contains only spin-polarized candidates, whereas S4 contains only unconventional candidates. The difference between S1/S3 and S2/S4 reflects two distinct prediction targets: S1 and S3 identify the more restrictive spin-polarized class with $d=1$, whereas S2 and S4 identify unconventional magnetism more broadly through $d>0$.

S5 represents a mixed candidate ensemble in which both conventional and unconventional magnetic states remain symmetry compatible. It should therefore not be interpreted as a negative prediction for unconventional magnetism. Rather, the crystallographic structure and magnetic-sublattice information alone do not provide sufficient selectivity, and additional information such as the magnetic-cell size, propagation vector, diffraction constraints or energetic calculations is required.

S6 represents the opposite limiting case: all compatible candidates within the present search space are conventional. This provides no positive indication of unconventional magnetism within the assumptions of the workflow, but it does not prove that the experimentally realized magnetic state must be conventional.

More generally, the predictive statements associated with S1--S6 are conditional on the candidate space considered here. Magnetic structures with $I_k>4$ are not included, and we further assume that the parent SG of the magnetic SSG is identical to the space group of the non-magnetic crystal. Magnetic orders that enlarge the magnetic cell beyond this range or lower the spatial symmetry therefore lie outside the present search space. Consequently, even when all retained candidates are conventional, a lower-symmetry magnetic state outside the present construction may still be unconventional. An S6 classification should therefore be understood as the absence of unconventional candidates within the present search space, rather than as a strict exclusion of unconventional magnetism.

Because of these limitations, the predictive value of the hierarchy for real materials cannot be established from the symmetry construction alone and must be tested against experimentally determined magnetic structures. The MAGNDATA benchmark presented below provides this validation. It shows that S1--S4 preferentially identify materials whose experimentally realized magnetic states are unconventional, whereas S5 and S6 exhibit weak or absent positive selectivity toward unconventional magnetism. The hierarchy should therefore be interpreted as a symmetry-based measure of material-level selectivity whose predictive significance is established empirically through the benchmark.

\subsection{Role of \texorpdfstring{$\Ik=1$}{Ik=1}}

The explicit use of the $I_k=1$ subset in S3 and S4 reflects an important practical feature of magnetic-structure determination. Although determining the full arrangement of magnetic moments can require detailed magnetic diffraction or spectroscopy, whether the magnetic order enlarges the crystallographic unit cell is often considerably easier to establish. In diffraction experiments, for example, the presence or absence of additional magnetic propagation vectors can directly constrain the magnetic-cell index.

Knowledge that $I_k=1$ therefore provides a simple additional filter on the symmetry-generated candidate ensemble without requiring the complete magnetic configuration. For a material whose full $C_{\leq4}$ set contains both conventional and unconventional states, restricting the candidates to $C_1$ may eliminate all conventional possibilities and lead to an S3 or S4 classification.

S3 and S4 should therefore not be regarded simply as weaker versions of S1 and S2. Rather, they represent predictions obtained after incorporating one additional experimentally accessible constraint. The present hierarchy emphasizes $I_k=1$ because magnetic structures without translational-cell enlargement are both common experimentally and straightforward to distinguish from larger-cell magnetic orders. The predictive value of this additional constraint is examined quantitatively in the MAGNDATA benchmark below.

%################################################################################
%##### SECTION: Benchmark against experimentally determined magnetic structures
%################################################################################

\section{Benchmark against experimentally determined magnetic structures}

To evaluate whether the symmetry-generated candidate ensembles carry predictive information for real materials, we benchmark the workflow against experimentally determined magnetic structures collected from MAGNDATA. The experimentally resolved magnetic configuration is used only as the ground truth for evaluation, whereas the prediction is reconstructed from the corresponding non-magnetic crystallographic structure and magnetic-sublattice information using the same symmetry-based workflow described above.

The benchmark addresses two complementary questions. We first examine whether the experimentally realized SSG is contained in the symmetry-generated candidate ensemble. More importantly, we then test whether the experimentally realized magnetic character agrees with the material-level prediction inferred from the complete candidate ensemble.

\subsection{Construction of the MAGNDATA benchmark set}

We start from 1,626 experimentally determined magnetic structures collected from MAGNDATA. To construct the benchmark input without using the experimentally known moment configuration, we retain only the underlying crystallographic structure and the magnetic-sublattice information and apply the input conditions defined above. Structures with partial atomic occupancies, non-compensated magnetic order, more than one magnetic chemical species, more than one magnetic Wyckoff position, or only a subset of symmetry-equivalent sites carrying magnetic moments are excluded before the symmetry-generation procedure. After this initial screening, 868 structures remain as valid inputs to the present workflow.

The excluded entries reflect the scope of the present implementation. Of the 1,626 initial MAGNDATA structures, 100 cannot be converted into fully ordered crystallographic inputs, predominantly because of partial occupancies, leaving 1,526 ordered structures. Among these, 1,213 are experimentally antiferromagnetic and are therefore relevant to the compensated-magnetism benchmark considered here. Applying the magnetic-sublattice conditions further excludes 345 entries: 130 contain more than one magnetic chemical species, 155 contain magnetic atoms on more than one inequivalent Wyckoff position, and 60 contain only a subset of the relevant symmetry-equivalent sites carrying magnetic moments. This leaves 868 structures as valid inputs to the symmetry-generation procedure.

The symmetry-generation procedure is then applied to the remaining 868 structures. Of these, 862 admit at least one compatible zero-net-moment SSG candidate within the present search space, whereas six do not. These six structures cannot be represented within the present candidate space, which assumes an SSG parent SG identical to the non-magnetic space group and $I_k\leq4$. The final benchmark therefore contains 862 materials for which a complete symmetry-generated candidate ensemble can be constructed. This experimentally established dataset provides a sufficiently large reference set for quantitatively testing the predictive selectivity of the present approach.

\subsection{Recovery of experimental SSGs}

As a first test, we examine whether the experimentally realized SSG is contained in the symmetry-generated candidate ensemble. Among the 862 benchmark materials, the experimental SSG is recovered exactly for 571, corresponding to an exact-recovery rate of $571/862=66.24\%$.

Within the present construction, exact recovery primarily tests whether the experimentally realized magnetic order is compatible with the assumption that the SSG parent SG remains identical to the non-magnetic crystallographic space group, together with the other restrictions of the candidate search. The 66.24\% recovery therefore shows that this symmetry-preserving construction applies to a substantial fraction of experimentally known magnetic structures. Exact recovery of the magnetic SSG, however, is not the primary objective of the present work; the central question is whether the candidate ensemble correctly predicts the conventional or unconventional magnetic character of the experimentally realized state.

\subsection{S1--S6 benchmark}

We next evaluate the full S1--S6 hierarchy introduced above. Each of the 862 benchmark materials is assigned a prediction class using only its symmetry-generated candidate ensemble, while the experimentally realized SSG is classified independently as conventional, unconventional non-spin-polarized, or spin polarized.

The results are summarized in Table~\ref{tab:S1_S6_benchmark}.

\begin{table}[htbp]
    \centering
    \caption{Benchmark of the S1--S6 material-level prediction hierarchy against experimentally determined magnetic structures in MAGNDATA. Experimental states are classified as conventional, unconventional but non-spin-polarized, or spin polarized according to the reciprocal-space spin-texture dimensionality of the reported SSG.}
    \label{tab:S1_S6_benchmark}
    \begin{tabular}{ccccc}
        \hline
        Class & Total & Conventional & Unconventional non-SP & Spin polarized \\
        \hline
        S1 & 5   & 2   & 0  & 3  \\
        S2 & 52  & 9   & 29 & 14 \\
        S3 & 128 & 56  & 8  & 64 \\
        S4 & 93  & 23  & 24 & 46 \\
        S5 & 539 & 490 & 23 & 26 \\
        S6 & 45  & 43  & 0  & 2  \\
        \hline
    \end{tabular}
\end{table}

The benchmark reveals a clear separation between the prediction classes. In S2, 43 of 52 materials are classified as unconventional, corresponding to an unconventional fraction of 82.7\%. In S4, 70 of 93 materials are classified as unconventional, corresponding to 75.3\%. S3 targets the more restrictive spin-polarized class: 64 of 128 materials are classified as spin polarized, while another 8 are unconventional but non-spin-polarized.

By contrast, S5 and S6 are strongly dominated by states classified as conventional. Among S5 materials, 490 of 539 are classified as conventional, corresponding to 90.9\%. For S6, 43 of 45 are classified as conventional, corresponding to 95.6\%.

The S1 subset contains only five benchmark materials, of which three are classified as spin polarized. Because of this small sample size, its numerical fraction should not be interpreted independently of the broader trend across the hierarchy.

Taken together, the benchmark demonstrates that the S1--S6 hierarchy carries meaningful material-level selectivity. S1--S4 preferentially identify states classified as unconventional or spin polarized, whereas S5 and S6 are strongly enriched in conventional magnetism. This empirical separation validates the predictive significance of the hierarchy beyond the symmetry construction itself.

\subsection{Role of \texorpdfstring{$\Ik=1$}{Ik=1}}

The MAGNDATA benchmark also provides a direct test of the additional predictive information carried by the magnetic-cell index. We focus on materials for which the experimentally realized structure has $I_k=1$ and examine the corresponding $I_k=1$ candidate ensemble.

Among these materials, 161 have a nonempty $I_k=1$ candidate set containing only unconventional states: 3 belong to S1, 40 to S2, 56 to S3 and 62 to S4. Independent classification of their experimental SSGs identifies all 161 experimental states as unconventional. Thus, the prediction agrees with the experimental classification for all 161 benchmark cases satisfying both conditions.

This result provides a particularly clear validation of the logic underlying S3 and S4. Even when the full $I_k\leq4$ candidate ensemble is mixed, knowledge that the experimentally realized magnetic structure has $I_k=1$ can eliminate the remaining conventional possibilities and yield a highly selective prediction of unconventional magnetic character.

Among 523 experimentally reported structures with $I_k=1$, 380 satisfy the present structural and compatibility conditions and admit at least one compatible $I_k=1$ candidate. The experimental SSG is recovered exactly for 348 of these 380 structures, corresponding to an exact-recovery rate of 91.58\%, compared with 66.24\% for the full benchmark set. Separately, the 161/161 result demonstrates the selectivity of the unconventional classification for entries with experimental $I_k=1$ whose generated $I_k=1$ candidates are all unconventional.

%################################################################################
%##### SECTION: High-throughput search in the Materials Project
%################################################################################

\section{High-throughput search in the Materials Project}

Having established the predictive selectivity of the candidate-ensemble approach using experimentally determined magnetic structures, we next apply the same workflow on a much larger scale to magnetic materials collected from the Materials Project. In contrast to MAGNDATA, where experimentally determined magnetic structures are available, the Materials Project is used here only to provide crystallographic structures and magnetic-sublattice information. The objective is therefore to identify materials whose symmetry-compatible candidate ensembles strongly constrain unconventional magnetic character without assuming a known magnetic ground state.

\subsection{Materials Project input and filtering}

We start from 58,996 Materials Project entries labelled as ferromagnetic or antiferromagnetic, including 55,594 FM and 3,402 AFM entries. Nonmagnetic and ferrimagnetic entries are not included in the input dataset. The magnetic configurations provided by the Materials Project are obtained from collinear first-principles calculations and are not taken here as definitive magnetic ground states. Instead, we use the site-resolved magnetic information only to identify the magnetic chemical species and the magnetic Wyckoff position that define the magnetic sublattice for the subsequent symmetry construction.

The same structural conditions introduced in the general workflow are then applied. We retain materials containing one magnetic chemical species occupying a single symmetry-equivalent Wyckoff position, with all sites belonging to this Wyckoff position included in the magnetic sublattice. The structures are then standardized before entering the SSG construction. After these filters, 23,849 structures remain.

The inclusion of both FM- and AFM-labelled Materials Project entries is therefore intentional. Their original collinear magnetic configurations are not used in the prediction. Once the magnetic sublattice has been identified, the subsequent workflow independently constructs the compensated magnetic states compatible with the crystallographic structure. For application to a real material, the additional physical requirement is that no substantial net magnetization is observed experimentally; materials established to be ferromagnetic or ferrimagnetic would therefore lie outside the direct target of the present compensated-magnetism classification.

\subsection{Candidate generation}

The 23,849 standardized structures are processed using the symmetry-generation procedure described above. For each material, candidate SSGs are first selected by requiring their parent SG to coincide with the non-magnetic crystallographic space group, $I_k\leq4$, and zero net magnetization. Each remaining SSG is then tested for compatibility with the specified magnetic Wyckoff position, and an explicit magnetic structure is constructed when the onsite symmetry constraints permit a nonzero magnetic moment.

Across the complete Materials Project search, 8,916,418 material-specific candidate SSGs are tested in this way. Of the 23,849 input structures, 23,428 admit at least one compatible zero-net-moment magnetic candidate within the present search space, whereas 421 admit none. The 23,428 retained materials give rise to a total of 825,243 compatible magnetic structures.

The number of compatible candidates varies substantially among materials because it depends on the crystallographic space group, the magnetic Wyckoff position and the SSGs compatible with these constraints. No energetic ranking is introduced during this procedure. All successfully constructed candidates are retained and classified according to their SSG symmetry, so that the final prediction is determined by the complete candidate ensemble rather than by a preselected magnetic configuration.

\subsection{S1--S6 distribution}

The 23,428 materials with at least one compatible candidate are classified according to the S1--S6 hierarchy defined above. The resulting distribution is summarized in Table~\ref{tab:mp_s1_s6_distribution}.

\begin{table}[htbp]
    \centering
    \caption{Distribution of the Materials Project materials among the S1--S6 material-level prediction classes.}
    \label{tab:mp_s1_s6_distribution}
    \begin{tabular}{ccc}
        \hline
        Class & Number of materials & Candidate-set character \\
        \hline
        S1 & 533    & All $I_k\leq4$ candidates spin polarized \\
        S2 & 1,236  & All $I_k\leq4$ candidates unconventional \\
        S3 & 3,126  & All $I_k=1$ candidates spin polarized \\
        S4 & 1,667  & All $I_k=1$ candidates unconventional \\
        S5 & 14,329 & Mixed conventional/unconventional candidates \\
        S6 & 2,537  & All $I_k\leq4$ candidates conventional \\
        \hline
    \end{tabular}
\end{table}

Among the full $I_k\leq4$ candidate ensembles, 1,769 materials belong to S1 or S2. For these materials, every compatible candidate is unconventional, so their unconventional magnetic character is strongly constrained without requiring additional information about the magnetic-cell size. Among them, the 533 S1 materials satisfy the stronger condition that every compatible candidate is spin polarized.

When the candidate set is restricted to $I_k=1$, a substantially larger group of materials becomes strongly constrained. The S3 and S4 classes contain 3,126 and 1,667 materials, respectively. Considering all classes together, 6,248 materials have only unconventional candidates within their $I_k=1$ candidate set, of which 3,345 have only spin-polarized candidates. These numbers illustrate the substantial additional selectivity provided by knowledge that the magnetic order does not enlarge the crystallographic primitive cell.

The majority of the retained materials, 14,329, belong to S5 and therefore admit both conventional and unconventional candidate states. For these materials, the crystallographic structure and magnetic-sublattice information alone do not provide sufficient selectivity to determine the unconventional magnetic character. By contrast, 2,537 materials belong to S6 and contain only conventional candidates within the present search space.

More broadly, 20,891 of the 23,428 retained materials possess at least one unconventional candidate, whereas the remaining 2,537 are the conventional-only S6 materials. The much smaller S1--S4 subsets are therefore the more relevant outcome of the high-throughput search: rather than merely identifying materials in which unconventional magnetism is symmetry allowed, they isolate materials for which the compatible candidate ensemble itself provides a strong constraint toward unconventional or spin-polarized magnetic character. These high-selectivity candidates form the primary target set for subsequent material-specific calculations and experimental investigation.

\subsection{Catalogue of light-element S1--S4 materials}

To present the prioritized materials from the full Materials Project search, the accompanying repository includes a PDF catalogue of all S1--S4 entries satisfying a light-element composition filter. The non-relativistic SSG description is most directly applicable when spin--orbit coupling is negligible on the relevant energy scale\cite{liu2022SSG}. We therefore select magnetic species from Ti, V, Cr, Mn, Fe, Co, Ni and Cu and require every constituent element to have atomic number $Z\leq36$. This composition filter prioritizes candidates for weak-SOC behaviour; it is not a material-specific determination of SOC strength and does not alter the S1--S6 classification.

The catalogue contains 2,752 materials: 178 in S1, 704 in S2, 1,356 in S3 and 514 in S4, with 110,870 compatible magnetic candidates in total. Each entry gives the Materials Project ID, chemical formula, magnetic species, parent space group, and candidate counts and spin-texture labels for $I_k=1,2,3,4$. Entries are grouped by their original Materials Project AFM or FM input labels, which indicate data provenance rather than experimentally established magnetic order. The PDF is available as a separate attachment alongside the code archive at the repository specified in Code availability.

%################################################################################
%##### SECTION: First-principles study of VGe3
%################################################################################
\section{First-principles study of \texorpdfstring{\ch{VGe3}}{VGe3}}

\subsection{Crystal structure and candidate ensemble}

\ch{VGe3} provides a compact example in which the material-level S-class prediction can be followed through to explicit first-principles calculations. The Materials Project entry considered here is mp-672337. The non-magnetic parent structure has space group $Pm\bar{3}n$ (No.~223), and the magnetic V atoms occupy a single symmetry-equivalent Wyckoff position of the parent crystal. This satisfies the single-sublattice condition used in the high-throughput workflow.

Within the present $I_k\leq4$ search space, the workflow finds only three compatible zero-net-moment magnetic candidates, summarized in Table~\ref{tab:vge3_candidates}. All three candidates have a one-dimensional spin texture in reciprocal space and are therefore spin-polarized unconventional magnetic states in the terminology used throughout this work. Consequently, \ch{VGe3} belongs to the S1 class: every compatible candidate in the complete $I_k\leq4$ ensemble is spin polarized. This makes \ch{VGe3} a particularly selective case study, because no conventional zero-net-moment candidate remains within the present search space.

\begin{table}[htbp]
    \centering
    \caption{SSG-compatible zero-net-moment candidates generated for \ch{VGe3} (mp-672337) within $I_k\leq4$. The spin-texture label specifies the leading symmetry-allowed wave order, or the lowest odd and even orders for a mixed-wave one-dimensional texture. International symbols are listed in Table~\ref{tab:ssg_international_symbols}.}
    \label{tab:vge3_candidates}
    \begin{tabular}{ccccc}
        \hline
        Candidate & SSG & $I_k$ & Real-space order & Spin texture \\
        \hline
        01 & $223.1.2.3.\mathrm{L}$ & 1 & collinear & $i$ wave \\
        02 & $223.2.4.1.\mathrm{P}$ & 2 & coplanar & $f$ wave \\
        03 & $223.2.4.4$ & 2 & noncoplanar & $f+i$ wave \\
        \hline
    \end{tabular}
\end{table}

The three candidates also illustrate the usefulness of the candidate-ensemble formulation. They differ in magnetic-cell index and real-space spin geometry, but share the same qualitative reciprocal-space character: all possess a finite one-dimensional spin polarization. The $I_k=1$ candidate is a collinear $i$-wave altermagnetic state. The two $I_k=2$ candidates are a coplanar $f$-wave state and a noncoplanar $f+i$-wave state, respectively. For the latter, SSG $223.2.4.4$ permits an odd component beginning at third order and an even component beginning at sixth order along the same polarization axis. The three candidates therefore illustrate even-, odd- and mixed-wave symmetry within a single S1 ensemble. These labels characterize the symmetry-generated configurations; the magnetic symmetry reached during self-consistency is examined separately below.

\subsection{Symmetry analysis}
The SSG classification determines the allowed form of the spin polarization before any electronic-structure calculation is performed. For the collinear candidate $223.1.2.3.\mathrm{L}$, the spin texture is one-dimensional. Choosing the spin quantization axis as $\hat{z}_s$, the lowest nonzero invariant obtained from the SSG constraints appears at sixth order in momentum:
\begin{equation}
    F^{(6)}(\mathbf{k})
    =
    k_y^2 k_z^4
    - k_y^4 k_z^2
    - k_x^2 k_z^4
    + k_x^2 k_y^4
    + k_x^4 k_z^2
    - k_x^4 k_y^2 .
    \label{eq:vge3_iwave_basis}
\end{equation}
For this spin-conserving collinear candidate, the basis describes the symmetry-allowed momentum dependence of the signed spin splitting; it should not be interpreted as the magnitude of the spin expectation value of an individual spin eigenstate.

We refer to this as an $i$-wave spin texture. The signed spin splitting transforms according to this basis. At momenta left invariant modulo a reciprocal lattice vector by a spin-flipping SSG operation, the symmetry-related opposite-spin states are degenerate.

For the coplanar candidate $223.2.4.1.\mathrm{P}$, the SSG has parent SG 223, pure-lattice subgroup with $H$ labelled by SG 196 in the SSG database, and a two-dimensional real-space spin representation. The momentum-space spin texture is nevertheless one-dimensional, with spin axis along $\hat{z}_s$ in the spin frame used for the band calculations. The lowest-order basis function is
\begin{equation}
    F^{(3)}(\mathbf{k})=k_x k_y k_z ,
    \label{eq:vge3_fwave_basis}
\end{equation}
which is the $f$-wave form used below to interpret the calculated spin-projected bands. Here $k_x$, $k_y$ and $k_z$ denote reciprocal coordinates of the unfolded parent cubic cell, not the reciprocal coordinates of the doubled magnetic cell used in the $I_k=2$ VASP calculations.

For the $I_k=2$ magnetic cell used for candidates 02 and 03, the real-space transformation from the parent cubic basis to the magnetic-cell basis can be written as
\begin{equation}
    M=
    \begin{pmatrix}
    0 & 1 & 1\\
    1 & 0 & 1\\
    1 & 1 & 0
    \end{pmatrix},
    \qquad
    |\det M|=2 .
    \label{eq:vge3_supercell_matrix}
\end{equation}
If $(K_1,K_2,K_3)$ are the reciprocal coordinates in the folded magnetic-cell basis, then the parent reciprocal coordinates are
\begin{equation}
    \begin{pmatrix}
    k_x\\
    k_y\\
    k_z
    \end{pmatrix}
    =
    M^{-T}
    \begin{pmatrix}
    K_1\\
    K_2\\
    K_3
    \end{pmatrix}
    =
    \frac{1}{2}
    \begin{pmatrix}
    -1 & 1 & 1\\
    1 & -1 & 1\\
    1 & 1 & -1
    \end{pmatrix}
    \begin{pmatrix}
    K_1\\
    K_2\\
    K_3
    \end{pmatrix}.
    \label{eq:vge3_folded_coordinate_transform}
\end{equation}
Equivalently,
\begin{equation}
    2k_x=-K_1+K_2+K_3,\quad
    2k_y=K_1-K_2+K_3,\quad
    2k_z=K_1+K_2-K_3 .
    \label{eq:vge3_folded_coordinates}
\end{equation}
The ordinary nodal planes of Eq.~\eqref{eq:vge3_fwave_basis} are $k_x=0$, $k_y=0$ and $k_z=0$. In the folded Brillouin zone, the same zero-spin constraints may also occur on zone-boundary planes, because a little-group operation can map $\mathbf{k}$ back to the same crystal momentum modulo a reciprocal lattice vector while reversing $\hat{z}_s$. Thus the folded-zone zero-spin conditions can be written as
\begin{equation}
    -K_1+K_2+K_3\in\mathbb{Z},\quad
    K_1-K_2+K_3\in\mathbb{Z},\quad
    K_1+K_2-K_3\in\mathbb{Z}.
    \label{eq:vge3_folded_zero_spin_conditions}
\end{equation}

The folded band path used for the $I_k=2$ calculations is
\begin{equation}
    \Gamma\rightarrow X\rightarrow M\rightarrow R\rightarrow\Gamma ,
    \label{eq:vge3_band_path}
\end{equation}
with endpoints
\begin{equation}
    \Gamma=(0,0,0),\quad
    X=(0,\tfrac12,\tfrac12),\quad
    M=(\tfrac12,\tfrac12,1),\quad
    R=(1,1,1)
    \label{eq:vge3_folded_endpoints}
\end{equation}
in the folded reciprocal basis. Under Eq.~\eqref{eq:vge3_folded_coordinate_transform}, the four path segments correspond to the parent-cell paths listed in Table~\ref{tab:vge3_kpath_symmetry}. The first three segments satisfy at least one of the folded-zone zero-spin conditions in Eq.~\eqref{eq:vge3_folded_zero_spin_conditions} and obey the zero-polarization constraint specified below. The interior of the $R\rightarrow\Gamma$ segment is not protected in this way and is the part of the plotted path where the $f$-wave spin splitting is allowed.

\begin{table}[htbp]
    \centering
    \caption{Symmetry interpretation of the folded $I_k=2$ band path for the \ch{VGe3} $f$-wave candidate. The parent-cell coordinates are obtained from Eq.~\eqref{eq:vge3_folded_coordinate_transform}.}
    \label{tab:vge3_kpath_symmetry}
    \begin{tabular}{cccc}
        \hline
        Segment & Folded-cell path & Parent-cell path & $\sigma_z$ splitting \\
        \hline
        $\Gamma\rightarrow X$ & $(0,t,t)$ & $(t,0,0)$ & forbidden \\
        $X\rightarrow M$ & $(t,\tfrac12,\tfrac12+t)$ & $(\tfrac12,t,0)$ & forbidden \\
        $M\rightarrow R$ & $(\tfrac12+t,\tfrac12+t,1)$ & $(\tfrac12,\tfrac12,t)$ & forbidden \\
        $R\rightarrow\Gamma$ & $(q,q,q)$ & $(\tfrac{q}{2},\tfrac{q}{2},\tfrac{q}{2})$ & allowed away from endpoints \\
        \hline
    \end{tabular}
\end{table}

For completeness, one can express the protection criterion directly in terms of SSG operations. If an SSG operation
\begin{equation}
    g=\{U\Vert R|\boldsymbol{\tau}\}
    \label{eq:vge3_protecting_operation}
\end{equation}
satisfies
\begin{equation}
    R\mathbf{k}=\mathbf{k}+\mathbf{G}
    \label{eq:vge3_little_group_condition}
\end{equation}
and its effective spin action reverses the one-dimensional spin axis,
\begin{equation}
    U_{\mathrm{eff}}\hat{z}_s=-\hat{z}_s ,
    \label{eq:vge3_spin_axis_reversal}
\end{equation}
then a nondegenerate Bloch state satisfies $S_z(\mathbf{k})=-S_z(\mathbf{k})$, giving $S_z(\mathbf{k})=0$. For a degenerate multiplet forming a complete symmetry-invariant subspace, the corresponding constraint is a vanishing spin trace over the multiplet. On the three protected segments in Table~\ref{tab:vge3_kpath_symmetry}, representative operations can be chosen as follows, written in parent cubic reciprocal coordinates. On $\Gamma\rightarrow X$, the operation
\begin{equation}
    R_{\Gamma X}=
    \begin{pmatrix}
    1 & 0 & 0\\
    0 & -1 & 0\\
    0 & 0 & 1
    \end{pmatrix},
    \qquad
    U_{\mathrm{eff},\Gamma X}=
    \begin{pmatrix}
    1 & 0 & 0\\
    0 & -1 & 0\\
    0 & 0 & -1
    \end{pmatrix}
    \label{eq:vge3_gamma_x_operation}
\end{equation}
leaves $(t,0,0)$ invariant and reverses $\hat{z}_s$. On $X\rightarrow M$, the operation
\begin{equation}
    R_{XM}=
    \begin{pmatrix}
    1 & 0 & 0\\
    0 & 1 & 0\\
    0 & 0 & -1
    \end{pmatrix},
    \qquad
    U_{\mathrm{eff},XM}=
    \begin{pmatrix}
    1 & 0 & 0\\
    0 & -1 & 0\\
    0 & 0 & -1
    \end{pmatrix}
    \label{eq:vge3_x_m_operation}
\end{equation}
leaves $(\tfrac12,t,0)$ invariant and reverses $\hat{z}_s$. On $M\rightarrow R$, the relevant protection is a folded-zone-boundary little-group condition. One representative operation is
\begin{equation}
    R_{MR}=
    \begin{pmatrix}
    0 & 1 & 0\\
    -1 & 0 & 0\\
    0 & 0 & 1
    \end{pmatrix},
    \qquad
    U_{\mathrm{eff},MR}=
    \begin{pmatrix}
    0 & 1 & 0\\
    1 & 0 & 0\\
    0 & 0 & -1
    \end{pmatrix},
    \qquad
    \boldsymbol{\tau}=(\tfrac12,\tfrac12,\tfrac12).
    \label{eq:vge3_m_r_operation}
\end{equation}
It maps
\begin{equation}
    R_{MR}(\tfrac12,\tfrac12,t)
    =
    (\tfrac12,-\tfrac12,t)
    =
    (\tfrac12,\tfrac12,t)+(0,-1,0),
    \label{eq:vge3_m_r_mapping}
\end{equation}
and therefore returns the momentum to the same point in the folded Brillouin zone while reversing the spin-texture axis. This imposes zero polarization for nondegenerate states and a vanishing spin trace for symmetry-invariant degenerate multiplets on $M\rightarrow R$, even though the parent-cell polynomial $k_xk_yk_z$ is nonzero on $(\tfrac12,\tfrac12,t)$.

\subsection{Computational details}
First-principles calculations were performed using VASP 6.4.1 with the PBE exchange--correlation functional and PAW pseudopotentials. The representative calculations discussed here use DFT+$U$ with $U=3$~eV on the V $3d$ orbitals, $J=0$, $L_{\mathrm{MAXMIX}}=4$, a plane-wave cutoff of 500~eV and Gaussian smearing with $\sigma=0.05$~eV. Atomic positions and lattice vectors were fixed to the symmetry-generated structures; the calculations therefore compare magnetic states within the same crystallographic geometry rather than performing a full structural relaxation. The self-consistent calculations used a $\Gamma$-centered $9\times9\times9$ mesh in the corresponding magnetic cell. Spin--orbit coupling was not included, so the computed spin polarization should be compared with the non-relativistic SSG prediction.

The collinear $223.1.2.3.\mathrm{L}$ state was calculated using a collinear VASP run. For direct energy comparison with the two $I_k=2$ states, the $I_k=1$ magnetic pattern was also represented in a trivial doubled cell containing four V and twelve Ge atoms. The two $I_k=2$ candidates were calculated using noncollinear VASP runs. The coplanar candidate was initialized with the first four V moments proportional to
\begin{equation}
    (0,1,0),\quad
    (0,-1,0),\quad
    (1,0,0),\quad
    (-1,0,0),
    \label{eq:vge3_coplanar_initial_moments}
\end{equation}
and the Ge moments initialized to zero. The noncoplanar candidate was initialized with the V moments proportional to
\begin{equation}
    \begin{aligned}
    &(0,0.811176,0.584802),\quad
    (0,-0.811176,0.584802),\\
    &(0.811176,0,-0.584802),\quad
    (-0.811176,0,-0.584802).
    \end{aligned}
    \label{eq:vge3_noncoplanar_initial_moments}
\end{equation}

The energy comparison in Table~\ref{tab:vge3_energies} uses the same standard-valence PBE PAW pseudopotentials for V and Ge in all three calculations. The coplanar $f$-wave state is lower in energy than the collinear $i$-wave state by $825.94$~meV per 16-atom cell, or $206.48$~meV per formula unit. The unconstrained noncoplanar initialization relaxes into the coplanar $f$-wave state, indicating an energetic preference for coplanar order within this computational setup.

\begin{table}[htbp]
    \centering
    \caption{Representative DFT+$U$ energies for the \ch{VGe3} candidate states at $U=3$~eV, using the same standard PAW datasets for all entries. Energies are reported as $E_{\sigma\rightarrow0}$ in the 16-atom comparison cell. The labels identify the initial configurations; the coplanar and noncoplanar initializations converge to the same coplanar state within numerical precision.}
    \label{tab:vge3_energies}
    \begin{tabular}{ccccc}
        \hline
        Candidate & Initialization & Initial SSG & $E_{\sigma\rightarrow0}$ (eV) & $\Delta E$ (meV/f.u.) \\
        \hline
        01 & collinear $i$ wave & $223.1.2.3.\mathrm{L}$ & $-75.18658947$ & $206.48$ \\
        02 & coplanar $f$ wave & $223.2.4.1.\mathrm{P}$ & $-76.01252727$ & $0$ \\
        03 & noncoplanar $f+i$ wave & $223.2.4.4$ & $-76.01252602$ & $0.00031$ \\
        \hline
    \end{tabular}
\end{table}

The final local magnetic moments show that the low-energy solution remains magnetic. In the coplanar $f$-wave calculation, the four V moments converge to approximately
\begin{equation}
    (0,2.772,0),\quad
    (0,-2.772,0),\quad
    (2.772,0,0),\quad
    (-2.772,0,0)\ \mu_{\mathrm B},
    \label{eq:vge3_coplanar_final_moments}
\end{equation}
with a vanishing total moment and negligible induced moments on Ge.

We also carried out additional checks with alternative PAW choices and with $U$ varied between 0 and 5~eV. Those calculations are useful for assessing the robustness of the energetic tendency, but total energies from different PAW datasets cannot be subtracted directly because their energy reference zeros and valence-electron counts differ. Within each internally consistent PAW branch, the coplanar $f$-wave solution remains below the collinear $i$-wave solution over the tested $U$ range. This supports the use of the coplanar state as the representative low-energy S1 realization, while the main conclusion that all \ch{VGe3} candidates are spin-polarized follows from SSG symmetry and does not depend on the energetic details.

To examine the electronic structure of the noncoplanar candidate, we applied a penalty functional that constrains the local-moment directions while allowing their magnitudes to relax. The configuration used for main-text Fig.~2d was initialized from the converged coplanar charge density and canted by $10^\circ$ toward the noncoplanar direction, with V moments proportional to
\begin{equation}
    \begin{aligned}
    &(0,\cos10^\circ,\sin10^\circ),\quad
    (0,-\cos10^\circ,\sin10^\circ),\\
    &(\cos10^\circ,0,-\sin10^\circ),\quad
    (-\cos10^\circ,0,-\sin10^\circ).
    \end{aligned}
    \label{eq:vge3_canting10_directions}
\end{equation}
A penalty strength of $\lambda=10$ in the VASP convention was used. The resulting constrained configuration provides the noncoplanar bands in main-text Fig.~2d and is excluded from the unconstrained energy comparison.

\subsection{Band structures}
The band structures in main-text Fig.~2b--d compare the three candidates along the same physical momentum path. The computational and spin-projection procedures are given in the main-text Methods.

The collinear $i$-wave state exhibits splitting between conserved spin channels. For the coplanar $f$-wave state, SSG symmetry enforces zero spin polarization on $\Gamma\rightarrow X$, $X\rightarrow M$ and $M\rightarrow R$, with the constraint on degenerate states understood as a vanishing spin trace as described above. Nonzero polarization is allowed and observed along the interior of $R\rightarrow\Gamma$.

The constrained noncoplanar configuration retains the polarization on $R\rightarrow\Gamma$ and develops additional spin-polarized splitting on $X\rightarrow M$. These features connect its mixed $f+i$-wave response to the two limiting states, as discussed in Section~\ref{sec:mixed_wave_magnetism}.

%################################################################################
%##### SECTION: Mixed-wave magnetism from noncoplanar order
%################################################################################

\section{Mixed-wave magnetism from noncoplanar order}
\label{sec:mixed_wave_magnetism}

Mixed-wave magnetism combines odd and even momentum dependences within a one-dimensional spin texture. Its origin can be understood by separating two different symmetry constraints: the restriction of the spin polarization to a fixed axis and the restriction of its parity under momentum reversal. The first can remain present even when the second is absent. Here we explain this distinction and use the three symmetry-compatible candidates of \ch{VGe3} to connect mixed-wave magnetism with collinear altermagnetism and coplanar odd-wave magnetism. We then show how the nodal planes of the two limiting states intersect to give the protected nodal lines of the mixed-wave state.

\subsection{Spin polarization without a fixed parity}

For a one-dimensional texture, the generic-momentum SSG constraints restrict the polarization to $\mathbf S(\mathbf k)=s(\mathbf k)\hat{\mathbf n}$. This restriction follows from operations acting locally at generic momentum, whose spin parts have a one-dimensional common invariant subspace. It does not by itself determine how $s(\mathbf k)$ transforms between $\mathbf k$ and $-\mathbf k$.

To express the separate parity condition, consider the scalar transformation law in Eq.~\eqref{eq:scalar_spin_transformation} and denote the effective momentum action by $D_g=\det(U)R$. An operation satisfying
\begin{equation}
    D_g=-I,\qquad U\hat{\mathbf n}=\eta_g\hat{\mathbf n},
    \qquad \eta_g=\pm1,
    \label{eq:mixed_parity_operation}
\end{equation}
imposes $s(\mathbf k)=\eta_gs(-\mathbf k)$ at every momentum. The polarization is therefore even for $\eta_g=+1$ and odd for $\eta_g=-1$. In collinear and coplanar spin-polarized states, the spin-only symmetries provide the parity constraints discussed above. A noncoplanar magnetic configuration has no corresponding nontrivial spin-only symmetry fixing the parity, although its combined spin-space operations can still impose such a constraint. Noncoplanarity alone therefore does not imply mixed-wave symmetry.

Mixed-wave symmetry becomes possible when no operation fixes a single parity, while the remaining SSG operations still enforce a one-dimensional spin subspace. The odd and even polynomial constraints must then be examined separately. When both sectors admit nonzero basis functions, the same scalar polarization allows both contributions,
\begin{equation}
    s(\mathbf k)=s_{\mathrm{odd}}(\mathbf k)+s_{\mathrm{even}}(\mathbf k),
    \qquad
    s(-\mathbf k)=-s_{\mathrm{odd}}(\mathbf k)+s_{\mathrm{even}}(\mathbf k).
    \label{eq:mixed_no_fixed_parity}
\end{equation}
If both contributions are present, the full texture is neither even nor odd. This loss of a definite parity is compatible with a fixed polarization axis and vanishing net magnetization. The symmetry classification specifies the allowed sectors; the electronic structure determines their amplitudes.

Related mixed-parity spin splitting has been proposed in collinear spin-orbital magnets\cite{Zhuang2026MixedParityCollinear}. Building on the spin-orbital altermagnetism framework\cite{Wang2025SpinOrbitalAltermagnetism}, that proposal realizes mixed parity through circularly polarized light in the presence of a staggered sublattice potential. For the noncoplanar candidates considered here, the SSG of the magnetic configuration itself constrains the spin polarization to a fixed axis while allowing both odd and even momentum dependences. This provides a route to mixed-wave magnetism without invoking additional spin-orbital order or external driving.

\subsection{Connecting altermagnetic and coplanar odd-wave order in \texorpdfstring{\ch{VGe3}}{VGe3}}

The \ch{VGe3} candidates connect the two parity sectors through a simple canting of the real-space moments. We measure the canting angle $\theta$ from the $xy$ spin plane, so that $\theta=0$ is the coplanar limit and $\theta=\pi/2$ is the perpendicular collinear limit. For the four V sites in the doubled magnetic cell, the moments can be written as
\begin{equation}
    \begin{aligned}
    \mathbf m_1&=m(0,\cos\theta,\sin\theta),&
    \mathbf m_2&=m(0,-\cos\theta,\sin\theta),\\
    \mathbf m_3&=m(\cos\theta,0,-\sin\theta),&
    \mathbf m_4&=m(-\cos\theta,0,-\sin\theta).
    \end{aligned}
    \label{eq:mixed_vge3_canting}
\end{equation}
The moments have equal magnitude and sum to zero for every $\theta$. At $\theta=0$, they reproduce the coplanar $223.2.4.1.\mathrm{P}$ candidate, with odd $f$-wave polarization perpendicular to the spin plane. At $\theta=\pi/2$, only the compensated collinear pattern $(+,+,-,-)$ along $\hat{\mathbf z}_s$ remains. The first two V sites then become equivalent under the parent-cell translations, as do the last two. Reduction to the parent primitive cell gives the collinear $223.1.2.3.\mathrm{L}$ candidate, with even $i$-wave altermagnetic symmetry.

At a generic intermediate angle, the moments contain both the coplanar pattern and a staggered perpendicular component. This noncoplanar configuration realizes $223.2.4.4$. Its SSG retains a fixed reciprocal-space polarization axis $\hat{\mathbf z}_s$ and allows both the odd $f$-wave and even $i$-wave sectors. Their lowest allowed degrees are 3 and 6, with leading basis functions in parent cubic momentum coordinates
\begin{equation}
    \begin{aligned}
    F^{(3)}(\mathbf k)&=k_xk_yk_z,\\
    F^{(6)}(\mathbf k)&=(k_x^2-k_y^2)(k_y^2-k_z^2)(k_z^2-k_x^2).
    \end{aligned}
    \label{eq:mixed_vge3_basis}
\end{equation}
For each magnetic configuration specified by $\theta$, the electronic structure sets the amplitudes of these allowed contributions. Denoting them by $A_f(\theta)$ and $A_i(\theta)$, the leading texture is
\begin{equation}
    \mathbf S(\mathbf k;\theta)
    =\left[A_f(\theta)F^{(3)}(\mathbf k)
    +A_i(\theta)F^{(6)}(\mathbf k)+\cdots\right]\hat{\mathbf z}_s.
    \label{eq:mixed_vge3_texture}
\end{equation}
The additive $f+i$ label refers to these two real contributions to $S_z$; $i$ denotes the sixth-degree wave order. Here $\theta$ labels the real-space magnetic configuration, and the omitted terms are higher-order contributions in the two parity sectors. The coplanar endpoint admits only odd terms and the collinear endpoint only even terms, so
\begin{equation}
    A_i(0)=0,\qquad A_f(\pi/2)=0.
    \label{eq:mixed_vge3_endpoints}
\end{equation}
The same parity restrictions apply to the higher-order terms. Between these limits, both sectors are symmetry-allowed. When both are present, the texture combines the odd-wave character of the coplanar state and the even-wave character of the altermagnetic state along the same spin axis. This coexistence is the sense in which the two limiting states contribute to mixed-wave magnetism.

Symmetry does not determine the amplitudes or require them to follow $\cos\theta$ and $\sin\theta$; the electronic response must be calculated for the combined magnetic configuration. The canting construction relates the symmetries of the three candidates and does not establish their energetic stability. In the representative unconstrained calculations discussed above, the noncoplanar initialization relaxes toward the coplanar solution.

\subsection{From nodal planes to protected nodal lines}

The leading basis functions make the change in the nodal structure explicit. In the pure coplanar $f$-wave limit, zero polarization occurs on three planes through $\Gamma$,
\begin{equation}
    F^{(3)}(\mathbf k)=0
    \quad\Longleftrightarrow\quad
    k_x=0\ \text{or}\ k_y=0\ \text{or}\ k_z=0.
    \label{eq:mixed_f_nodal_planes}
\end{equation}
In the collinear $i$-wave limit, the corresponding condition gives six planes,
\begin{equation}
    F^{(6)}(\mathbf k)=0
    \quad\Longleftrightarrow\quad
    k_x=\pm k_y\ \text{or}\ k_y=\pm k_z\ \text{or}\ k_z=\pm k_x.
    \label{eq:mixed_i_nodal_planes}
\end{equation}
For these factorized basis functions, the three and six linear nodal factors correspond to their third- and sixth-degree wave labels.

When the two sectors coexist, the scalar polarization is $s=A_fF^{(3)}+A_iF^{(6)}+\cdots$. A plane on which only $F^{(3)}$ vanishes can carry a nonzero $i$-wave contribution, and a plane on which only $F^{(6)}$ vanishes can carry a nonzero $f$-wave contribution. The zeros common to both leading contributions satisfy
\begin{equation}
    F^{(3)}(\mathbf k)=0,\qquad F^{(6)}(\mathbf k)=0.
    \label{eq:mixed_common_nodes}
\end{equation}
For example, within $k_x=0$, the second condition reduces to
\begin{equation}
    F^{(6)}(0,k_y,k_z)=-k_y^2k_z^2(k_y^2-k_z^2)=0,
    \label{eq:mixed_plane_lifting}
\end{equation}
leaving $k_y=0$, $k_z=0$ or $k_y=\pm k_z$ in that plane. Including the analogous intersections in the other two coordinate planes gives nine distinct lines through $\Gamma$: three along $\langle100\rangle$ and six along $\langle110\rangle$.

These common lines are protected by the mixed-state SSG, beyond the leading-order expansion. On each line, an operation of $223.2.4.4$ leaves the momentum fixed and reverses the polarization, requiring $s=-s=0$. The spin-reversing operations have one-dimensional fixed subspaces and no two-dimensional fixed subspace through $\Gamma$. The mixed state therefore retains protected nodal lines at the intersections of the odd- and even-wave nodal planes, while neither family of planes remains protected as a whole. It combines the two allowed contributions to spin polarization and has no definite momentum-reversal parity when both are nonzero.

The protected lines form part of the full zero set of the scalar polarization $s(\mathbf k)$. Symmetry-related regions of opposite sign must be separated by zeros of a continuous polarization field. For a nonzero real-analytic scalar field in three dimensions, such sign-separating zeros include two-dimensional strata; ordinary lines and isolated points alone cannot separate the positive and negative regions. At regular zeros, where $\nabla_{\mathbf k}s\neq0$, the zero set is locally a smooth two-dimensional surface. Continuity therefore requires separating zeros, while regularity determines their local smoothness; neither argument fixes their detailed shape or global number. In the mixed-wave state, cancellation between the odd and even contributions generally shifts these surfaces away from the nodal planes of the separate components, with their geometry determined by the amplitudes and higher-order terms. The nine protected lines remain fixed by the SSG, while the surrounding nodal surfaces can deform. Thus, the change from protected planes to protected lines describes the symmetry-fixed subset of the nodes, rather than a reduction of the entire zero set to one dimension. For the polarization field, these nodes denote zero spin polarization; their relation to electronic band degeneracies requires a separate analysis.

\section{Energetic screening of tetragonal \texorpdfstring{\ch{Fe2SiO4}}{Fe2SiO4}}
\label{sec:fe2sio4_screening}

The Materials Project structure mp-18816 has tetragonal parent space group $I4_1/amd$ (No.~141). All 15 constructed zero-net-moment candidates with $I_k\leq4$ are unconventional, placing this structure in S2. The ensemble contains one collinear $d$-wave state, five coplanar $p$-wave states, one coplanar $f$-wave state, five noncoplanar $p+d$-wave states, one noncoplanar $f+d$-wave state, one coplanar two-dimensional texture and one noncoplanar three-dimensional texture.

The six coplanar odd-wave candidates each have a noncoplanar mixed-wave counterpart. Using the candidate indices in Table~\ref{tab:fe2sio4_screening}, the pairs are 04/07, 05/06, 08/11, 09/10, 12/14 and 13/15. Mapping the generated Fe moments to common crystallographic sites shows that each mixed configuration is the sum of the corresponding coplanar pattern and a perpendicular staggered component with the site-sign pattern of the collinear $d$-wave candidate 01, up to a global spin rotation and component amplitudes. Five pairs therefore connect $p$-wave and $p+d$-wave symmetry, and the remaining pair connects $f$-wave and $f+d$-wave symmetry. This construction relates real-space magnetic configurations and their allowed parity sectors; it does not prescribe a linear combination of their separately calculated bands or energies.

The computational settings and magnetic-cell meshes are given in Methods. The final electronic-step energy changes were below $10^{-5}$~eV per Fe atom. Energies are normalized per Fe atom and referenced to the coplanar $f$-wave candidate with SSG 141.2.4.3.P.

\begin{table}[htbp]
    \centering
    \caption{Relative energies of the 15 \ch{Fe2SiO4} candidates shown in main-text Fig.~2e, normalized per Fe atom and referenced to the lowest-energy candidate. SSG labels refer to the constructed candidates. International symbols are listed in Table~\ref{tab:ssg_international_symbols}.}
    \label{tab:fe2sio4_screening}
    \begin{tabular}{clrc}
        \toprule
        Candidate & SSG & $\Delta E$ (meV/Fe) & Texture dimension \\
        \midrule
        01 & 141.1.2.6.L & 2.031 & 1 \\
        02 & 141.1.8.2.P & 4.361 & 2 \\
        03 & 141.1.8.34 & 4.379 & 3 \\
        04 & 141.2.4.1.P & 5.254 & 1 \\
        05 & 141.2.4.3.P & 0.000 & 1 \\
        06 & 141.2.4.12 & 2.355 & 1 \\
        07 & 141.2.4.2 & 4.152 & 1 \\
        08 & 141.3.2.1.P & 4.163 & 1 \\
        09 & 141.3.4.2.P & 7.905 & 1 \\
        10 & 141.3.4.11 & 7.097 & 1 \\
        11 & 141.3.4.7 & 3.866 & 1 \\
        12 & 141.4.4.1.P & 9.906 & 1 \\
        13 & 141.4.4.2.P & 3.626 & 1 \\
        14 & 141.4.4.3 & 8.056 & 1 \\
        15 & 141.4.4.7 & 3.324 & 1 \\
        \bottomrule
    \end{tabular}
\end{table}

The calculated local moments retain Fe moments of approximately $3.70\,\mu_{\mathrm B}$. Their input SSG operation sets were preserved within a $0.005\,\mu_{\mathrm B}$ local-moment tolerance; the largest raw Fe-moment operation residual was $0.004\,\mu_{\mathrm B}$. For candidates 03, 06, 07 and 12--15, the numerical SSG identifier did not terminate with a group number for either input or output. Their labels are the target SSG numbers, supported by matching the full operation sets after moment symmetrization, with a maximum single-site correction of $0.00224\,\mu_{\mathrm B}$. This symmetry check supports the candidate assignments independently of their energetic ordering.

%################################################################################
%##### SECTION: Scope and limitations
%################################################################################

% \section{Scope and limitations}

\end{document}